\documentclass{aa}  

\usepackage{graphicx}
\usepackage{multirow}
\usepackage{txfonts}
\begin{document}

   \title{$\tau$~Herculid meteor shower on night 30/31 May, 2022, and properties of the meteoroids}

   \author{P. Koten \inst{1}
		  \and
		  L. Shrben\'{y} \inst{1}
		  \and
		  P. Spurn\'{y} \inst{1}
		  \and
          J. Borovi\v{c}ka \inst{1}
          \and 
          R. \v{S}tork \inst{1}
          \and
          T. Henych \inst{1}
          \and
          V. Voj\'{a}\v{c}ek \inst{1}
          \and 
          Jan M\'{a}nek \inst{1}
          }

   \institute{Astronomical Institute, CAS, Fri\v{c}ova 298, 25165 Ond\v{r}ejov, Czech Republic\\
              \email{pavel.koten@asu.cas.cz}
              }

   \date{Received dd-mm-yyyy; accepted dd-mm-yyyy}

 
  \abstract
   {A $\tau$~Herculid meteor outburst or even a storm was predicted by several models to occur around 5~UT on 31~May, 2022 as a consequence of the break-up of comet 73P/Schwassmann-Wachmann~3 in 1995. The multi-instrument and multi-station experiment was carried-out within the Czech Republic to cover possible earlier activity of the shower between 21 and 1~UT on 30/31~May.}
   {The goal of the paper is to report meteor shower activity which occurred before the main peak and provide a comparison with the dynamical simulations of the stream evolution. Physical properties of the meteoroids are also studied.}
   {Multi-station observations using video and photographic cameras were used for calculation of the atmospheric trajectories and heliocentric orbits of the meteors. Their arrival times are used for determination of the shower activity profile. Physical properties of the meteoroids are evaluated using various criteria based on meteor heights. Evolution of spectra of three meteors are studied as well.}
   {This annual but poor meteor shower was active for the whole night many hours before the predicted peak. A comparison with dynamical models shows that a mix of older material ejected after 1900 and fresh particles originating from the 1995 comet fragmentation event was observed. Radiant positions of both groups of meteors were identified and found to be in good agreement with simulated radiants. Meteoroids with masses between 10~mg and 10~kg were recorded. The mass distribution index was slightly higher than 2. A study of the physical properties shows that the $\tau$~Herculid meteoroids belong to the most fragile particles observed ever, especially among higher masses of meteoroids. Exceptionally bright bolide observed during the dawn represents a challenge for the dynamical simulations as it is necessary to explain how to transfer a half metre body to the vicinity of the Earth at the same time as millimetre sized particles.}
   {}


   \titlerunning{Earlier activity of $\tau$~Herculid meteor shower}
   \authorrunning{P. Koten et al.}

   \maketitle
%

\section{Introduction}

Comet 73P/Schwassmann-Wachmann~3, which was discovered on May 2, 1930, experienced a significant increase of activity during its return to the central region of the Solar system in 1995. The radio observations showed an increase of OH production, followed by a splitting of the nucleus weeks later \citep{Crovisier1996}. Because the comet is a parent body of $\tau$~Herculid\footnote{061 TAH code in IAU MDC database}, these fragmentation events drew attention among the meteor dynamicists who model the meteoroid streams evolution. \citet{Wiegert2005} analysed the meteoroids released from the comet during the 1995 break-up as well as on previous perihelion passages back to 1801 and predicted detectable activity for 2022 and 2049. Their model showed that meteoroids released in 1892 and 1897 may reach the Earth on 30/31 May 2022. No activity connected with the 1995 break-up was expected.

Later, several other authors investigated possible connections between the 1995 comet fragmentation events and future meteor shower activity. \citet{Luthen2001}, \citet{Horii2008} and \citet{Rao2021} found that the material released during 1995 comet break-up would pass the Earth very closely on 31 May, 2022 and may cause an outburst or even a meteor storm. Main difference against standard models was in assumption of higher ejection velocities. Analyses of Jeremie Vaubaillon from early 2022 showed that 2.5 times higher than the Whipple model ejection velocities would bring the meteoroids close to the Earth\footnote{\url{https://www.imcce.fr/recherche/campagnes-observations/meteors/2022the}}. 

It was confirmed by \citet{Ye2022} who ran two different models but still received the same results. Using the ejection velocities three times higher than that from Whipple's model they were able to move the material ejected in 1995 to the vicinity of the Earth and to produce observable meteor activity. Finally, with the very first observational results available, they shifted the necessary speed even higher to 4x to 5x of the Whipple's values.

All the models predicted the peak of the activity to occur at about 5 UT on 31 May. This timing together with the position of the radiant made North America the most suitable area for the observations. During the early months of 2022 it became obvious that a number of teams are heading to the western states of the USA to observe the predicted event. So, the coverage of the event was ensured. This led us to the decision to stay in Europe and to cover possible earlier activity of the 1995 outburst and/or activity connected with older material originating from the end of 19th century. 

Shortly after the predicted peak, first messages were published confirming that the meteor shower really materialized and reached significant activity although not the storm level. \citet{Jenniskens2022} reported that the CAMS network\footnote{\url{http://cams.seti.org/}} measured 2244 $\tau$~Herculid orbits and the activity peaked on May 31 at 4$^{h}$42$^{m}\pm$25$^{m}$~UT. The GMN network\footnote{\url{https://globalmeteornetwork.org/}} observed 1396 multi-station $\tau$~Herculid meteors with an activity peak at 4$^{h}$15$^{m}$~UT \citep{Vida2022}. More than 14 thousand single-station $\tau$~Herculid meteors were observed by the network. The activity profile derived from the visual observations \citep{Rendtel2022} shows a broad maximum before midnight with a peak around 23~UT followed by descent. The width of this maximum was about 3~hours. The lowest activity was observed at about 0:30~UT. Later, the numbers started to rise again and the main peak is reported at 5$^{h}$5$^{m}\pm$5$^{m}$~UT. 

Recently, \citet{Egal2023} modelled the 2022 $\tau$-Herculid outburst and compared results with observed data. They concluded that the first peak of shower activity recorded by radio experiment between 15 and 19~UT on May 30 was caused by the material released between 1900 and 1947. The main peak, which occurred at 4-4:30~UT, originated indeed from the 1995 break-up of the parent comet.

In this paper we report multi-station and multi-instrument observation of the $\tau$~Herculid meteor shower from central Europe which occurred hours ahead of the predicted activity peak. The paper is organised as follows: Section \ref{observation} describes the observation strategy, used instruments and data processing, Section \ref{shower_activity} the meteor shower activity profile, mass distribution and flux, Section \ref{radiants} the radiants and orbits, Section \ref{properties} the physical properties of meteoroids and spectra, and, finally, Section \ref{discussion} summarises the results and discusses their implications. The geocentric radiants and the heliocentric orbits of $\tau$-Herculid meteors can be found in Appendix.

\section{Observations, instruments, data processing}
\label{observation}

Preparations of the observational campaign started several weeks ahead of the predicted date of meteor shower activity. Different variants of the observations were considered. In optimal case, the campaign was planned to be carried-out directly within the Czech part of the European Fireball Network (EN). The mobile variant was also prepared. Two moveable teams were ready to be dispatched somewhere to western or southern Europe depending on the weather forecast to establish double station video and photographic observations. 

Perfect forecast was provided by the team of the Czech Hydrometeorological Institute (CHMI) in Prague on a daily basis. The decision was made a day before the peak of activity. The forecast gave us relatively high chances for clear sky in central Europe. Therefore we decided to stay in the Czech Republic and to utilise the background of already established stations.

\subsection{Instrumentation}
\label{instrumentation}

\begin{figure}
  \resizebox{\hsize}{!}{\includegraphics{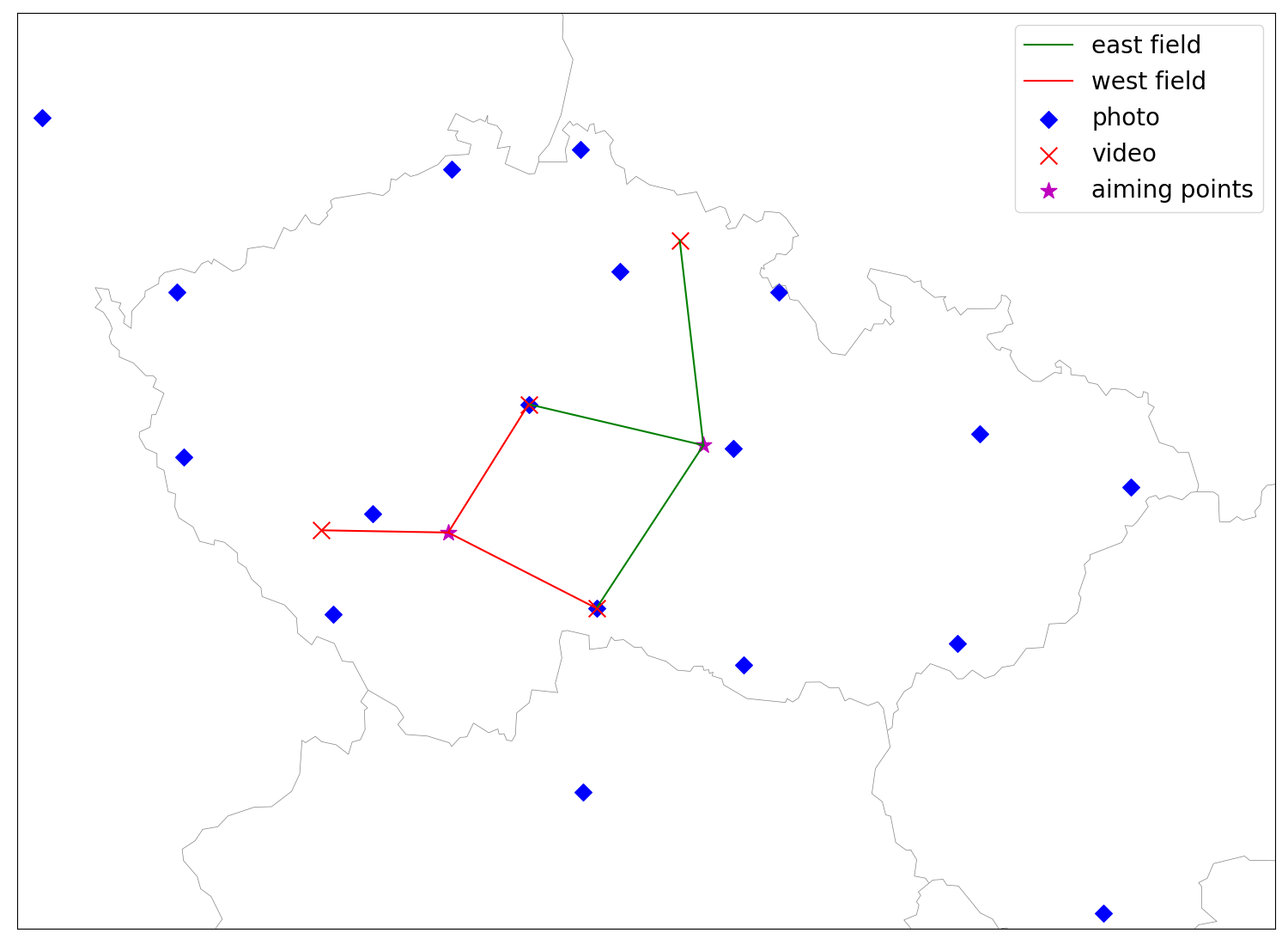}}
  \caption{Map of the Czech Republic showing the EN stations marked by blue diamonds, stations with video cameras (red crosses), and aiming points for video experiments by magenta stars.}
  \label{fig_mapCZ}
\end{figure}

Figure~\ref{fig_mapCZ} shows the EN stations in the part of central Europe together with locations of the video experiment cameras. All EN station are equipped with the high resolution digital autonomous observatories (DAFO), each of them consists of a pair of full-frame Canon 6D digital camera and Sigma fish-eye 8mm F3.5 lens and an electronic LCD shutter. Moreover, each DAFO is equipped with an all-sky radiometer with time resolution of 5000 samples per second. More details about DAFO can be found in \citet{Spurny2017}. 

Nearly half of EN stations are equipped also with the spectral version of the DAFO, called SDAFO \citep{Borovicka2019}. It differs from DAFO by using the 15mm F2.8 Sigma lens, lack of the LCD shutter, and presence of non-blazed plastic holographic gratings with 1000 grooves per mm in front of the lenses. Spectra can be obtained for fireballs brighter than about magnitude $-7$. Nevertheless, larger lens and lack of shutter provides higher sensitivity to meteor images than DAFO. SDAFO images were therefore used in this study also for computing meteor trajectories. However, they do not provide velocity and brightness information.

EN stations are also equipped with supplementary video arrays based on 4 megapixel Dahua IP cameras with resolution 2688x1520 pix and 20 or 25 frames per second in MJPEG encode mode. Each camera had 56$^{\circ}$ of the horizontal field of view. The Kun\v{z}ak and Ond\v{r}ejov stations host batteries of these cameras that cover the whole sky and also contain the holographic grating. Other EN stations are usually equipped by one or two Dahua cameras without grating, which are recording in H.265 encode mode. More details about these cameras can be found in \citet{Borovicka2019} or in \citet{Shrbeny2020}.

The video experiment was also carried-out inside the fireball network in order to observe faint meteors. Two three station experiments were established for the $\tau$~Herculid campaign. Routine automatic video observations are carried-out using Maia cameras on a daily basis as the double-station experiment at Kun\v{z}ak and Ond\v{r}ejov stations. The Maia camera consists of JAI CM-040 CCD camera, Pentax 1.8/50mm lens and Mullard XX1332 image intensifier and provides spatial resolution 776 x 582 pixels. The maximum frame rate is 61.15 per second, which was artificially lowered to half because of the very low $\tau$~Herculid meteors velocity. Detailed description of the system is in \citet{Koten2011}. The Maia cameras covered western field of the multi-station video experiment. The cameras aim at a fixed elevation of about 100~km above the surface. 

The eastern field was covered by the manually operated video cameras. This experiment is based on a digital DMK 23G445 GigE monochrome camera, again coupled with Mullard XX1332 image intensifier. It is connected with a long focal Canon 2.0/135 mm lens to be able to detect even fainter meteors. The CCD sensor provides spatial resolution 1280 x 960 pixels and time resolution 30 frames per second. More details are provided e.g. by \citet{Koten2020}. Because of the very low $\tau$~Herculid meteor velocity, the aiming point of the cameras was set at 80~km elevation. 

In the case of Maia cameras, the 50mm lenses provide a circular field-of-view of about 52$^{\circ}$. The DMK cameras with 135mm lenses provide a field-of-view of about 22$^{\circ}$. The limiting meteor magnitude is about +5.5$^{m}$ for Maia cameras and +7.0$^{m}$ for DMK cameras. A spectral video camera was operated at Ond\v{r}ejov station. Again, it used a DMK camera, XX1332 image intensifier and Jupiter 2.0/85mm lens with a blazed spectral grating with 600 grooves/mm. It was aimed at the eastern field.

Each double-station experiment was accompanied by another station using 4 Mpx Dahua cameras and mobile photographic Canon 6D camera. The western station at \v{S}tipoklasy supplemented the Maia cameras, the northern station at Hostinn\'{e} was complementary to the video DMK experiment. These two cameras are more sensitive, newer models than other IP cameras used within the network. In the case of very slow $\tau$~Herculid meteors they were able to detect meteors up to +4$^{m}$.

\subsection{Data processing}
\label{dataproc}

Each of the experiments employed in this campaign has its own way of data processing. For details, see the above mentioned papers. Generally, following the observations each data set was searched for the meteors. Found meteors were catalogued and their lists compiled together to obtain a global view of the recorded data and to be able to identify double or multi-station events. Such meteors were manually measured using FishScan software \citep{Borovicka2022} and their atmospheric trajectories and heliocentric orbits were computed by the Boltrack program using standard procedures \citep{Borovicka2022}. For the single station meteors a shower membership was estimated depending on their movement direction and angular velocity.

For meteors with known atmospheric trajectories it is possible to measure the absolute brightness at each point. Firstly, the calibration curve is constructed based on the measured stars with known brightness. Then, the measured signal of the meteor is transformed into the apparent brightness using such a curve. Finally, this value is recomputed on the distance of 100~km to obtain the absolute brightness. The photometric mass is computed using standard procedures \citep{Ceplecha1987}. The luminous efficiency according to \citet{Pecina1983} was used for the integration of the meteor light curve of video meteors, whereas the method of \citet{Revelle2001} was applied on the photographic records. In the case of photographic meteors where their photometry could not be done because there is no dynamic data, meteor photometry was determined from IP cameras. The IP cameras were
photometrically calibrated using photographic meteors, where the photometry was determined from DAFO by our standard procedures.

Only meteors with complete or almost complete light curves, i.e. whole recorded luminous trajectory, were used. Altogether, it counts 80 double or multi-station meteors -- 23 of them were recorded by the DMK cameras, 34 by the Maia cameras and 23 by the photographic cameras. Additional 55 single station video meteors were used for the mass distribution index computation.

\section{Meteor shower activity}
\label{shower_activity}

Note that the weather forecast was fulfilled quite well. Although the observational conditions were not ideal for some parts of the night because it was overcast especially at the beginning of the observational window at some of the stations, the experiment was finally carried-out successfully. The decision to stay within the fireball network and not to travel abroad was right. The fireball network cameras were operational for the whole night 30/31 May from dusk to dawn. The video experiment was planned from 20 to 1:30~UT. It started later in the south due to cloudy sky but after 22:30~UT all the cameras were recording under almost clear sky.

The shower membership was determined using the $D_{SH}$ criterion \citep{Southworth1963}. As the reference orbits, both the catalogue as well as the parent comet\footnote{\url{https://ssd.jpl.nasa.gov/}} orbits (solution K222/29 for the comet's 2017 apparition) were used. The meteors with $D_{SH} < 0.2$ were accepted as $\tau$~Herculids. In case of single station meteors, their trajectories were calculated under assumption that they belong to the meteor shower and if the radiant was found within 5~degrees from the mean radiant (compact group -- see Section~\ref{radiants}), they were also counted as $\tau$~Herculids.

\subsection{Activity profile}
\label{activity_profile}

To show the total activity profile, all the meteors identified as $\tau$~Herculids -- single and double-station -- detected at least by one of all used cameras were taken into account. We are aware that the numbers of meteors recorded by our cameras cannot be comparable with the numbers detected by the CAMS and GMN networks but still can provide us with an insight into the activity level. Altogether 207 $\tau$~Herculid meteors were counted in 30 minute intervals and these numbers were corrected on the zenith distance of the radiant. The resulting value is noted as the corrected hourly rate $cHR$ but this is not directly comparable with the zenith hourly rate traditionally used by visual observers. Due to low numbers of meteors shorter time intervals provide profiles containing a number of fluctuations which may or may not be important. 

\begin{figure}
  \resizebox{\hsize}{!}{\includegraphics{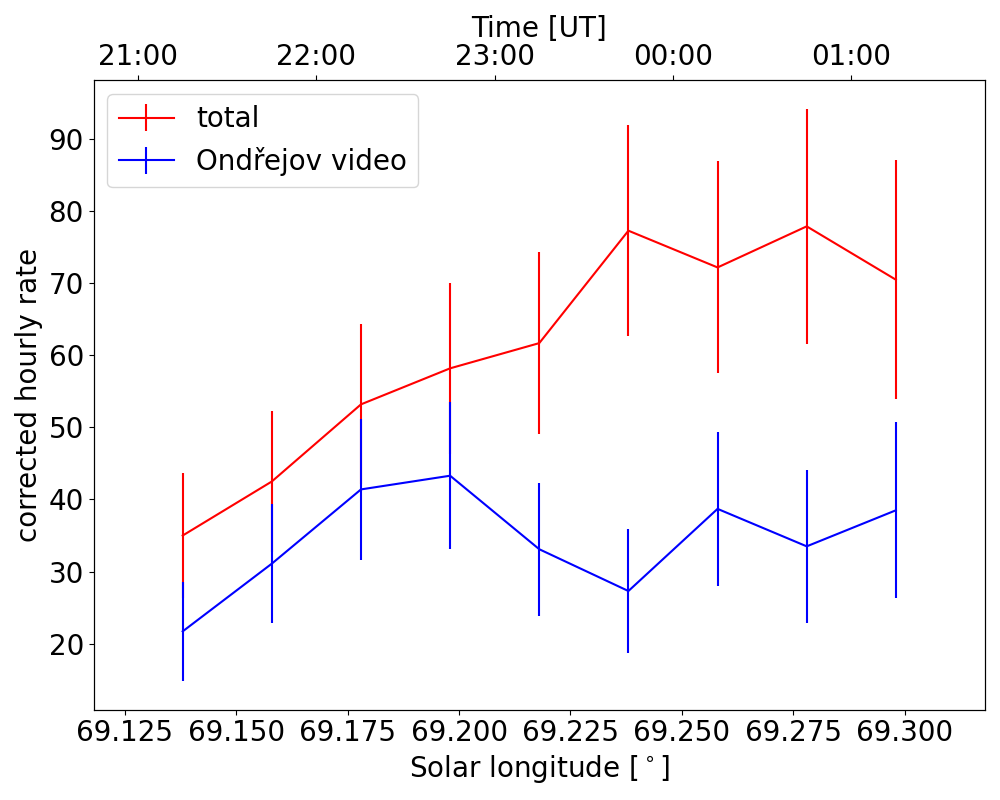}}
  \caption{An activity profile of $\tau$~Herculid meteor shower. The red line represents a profile which is based on 207 meteors detected by all instruments and counted in 30 minute intervals. A profile calculated only from Ond\v{r}ejov video camera's data is drawn in blue. The edge intervals are not included due to incompleteness of data (clouds at the beginning, dawn at the end).}
  \label{fig_activity}
\end{figure}

According to the profile in Figure~\ref{fig_activity} the activity started to increase after 21~UT and reached a relatively flat maximum around 23:30~UT. The activity was fluctuating between 70 and 80 meteors per hour for almost two hours. It began to decrease again after 1~UT mainly due to the beginning of the dawn. 

As the video meteors dominate the $\tau$~Herculid meteor sample and two of four video cameras were disabled due to cloudy sky before 23~UT, this profile (we call it a 'total' because all instruments are included) may be biassed. To eliminate this bias, we concentrated only at Ond\v{r}ejov station data, because the sky became clear here much earlier. Only $\tau$~Herculid meteors recorded by the DMK and Maia cameras located at this station were used for the profile calculation in the same way the above mentioned profile was. The result is depicted by the blue line in Figure~\ref{fig_activity}.

The comparison shows that the maximum activity occurred earlier between 22 and 23~UT. The apparent later increase of the activity on the 'total' graph was probably a consequence of the fact that more video-cameras contributed the counts. There are some fluctuations seen after 23:30~UT but due to the low number of the meteors it is difficult to say if they are significant or not. We can only say that relatively high activity continued till 1~UT. Such a result is in better agreement with the visual observation activity profile presented in \citet{Rendtel2022}. Initially, this timing suggested that an encounter with older filaments from the end of 19$^{th}$ century could be observed.

\subsection{Mass distribution}
\label{masses}

\begin{figure}
  \resizebox{\hsize}{!}{\includegraphics{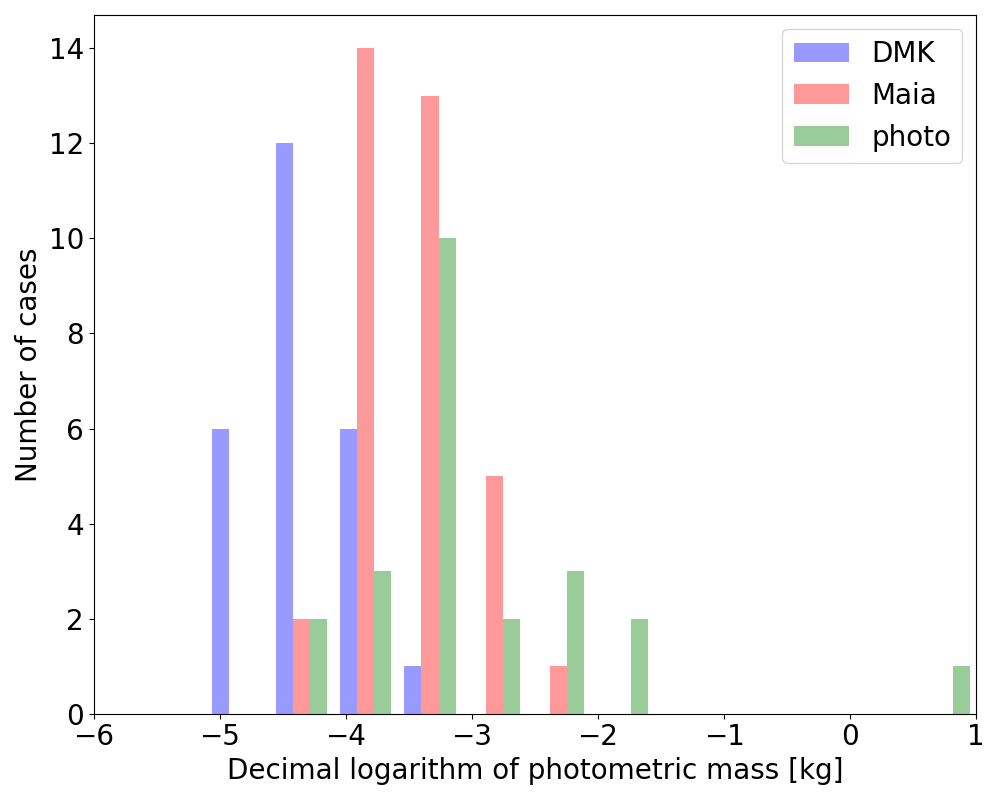}}
  \caption{Histogram of photometric masses of meteoroids as recorded by different kinds of instruments.}
  \label{fig_mass_hist}
\end{figure}

Due to different kinds of instruments used, the range of the photometric masses extends from 10$^{-5}$ to about 10~kg. As expected, the faintest meteors (the lightest meteoroids) were recorded by DMK video cameras, the brightest ones (the heaviest) by the photographic cameras (Figure~\ref{fig_mass_hist}).

As was mentioned above, the already derived corrected hourly rate is instrument dependent and cannot be directly compared with similar quantities either visual or instrumental. Therefore, it is necessary to convert the hourly rate to some kind of universal quantity, which does not depend on instrument parameters. Such a quantity is a flux of meteoroids. To be able to derive it, we need to determine the mass distribution index of the meteor shower. Usually, it is measured as a linear part of the best fit of log-log cumulative distribution (logarithm of mass vs. logarithm of number of meteors). If the slope of the fit is $k$ then the mass distribution index $s = 1 - k$. If $s < 2$ then more mass is in larger particles whereas $s > 2$ indicates that the stream contains many small particles \citep{Blaauw2011}.

The observations were carried-out by two very different kinds of instruments. Firstly, there were two double station video experiments using cameras with limited size of field-of-view, secondly the photographic cameras of the fireball network are the all-sky cameras distributed in vast areas of central Europe (see Figure~\ref{fig_mapCZ}). It means that the collection area of both experiments is incomparable. Taking into account all the masses of observed meteors would provide biassed results of the mass distribution index. Moreover, the observation conditions were changing at the beginning of the night, which is relatively simple to describe for the double station experiment but more difficult for the all sky cameras scattered around a bigger area. For all these reasons, only video data was used for the mass distribution index determination. To correct for different collection areas, the numbers of meteors detected by DMK cameras were multiplied by the ratio of collection areas of both instruments (which is approximately 8). The sample was extended by adding the single station meteors, for which the maximum apparent brightness was measured. Using the relation maximum apparent brightness vs. photometric mass, which was found using the double station meteors, the photometric mass of the single station meteors was estimated. 

\begin{figure}
  \resizebox{\hsize}{!}{\includegraphics{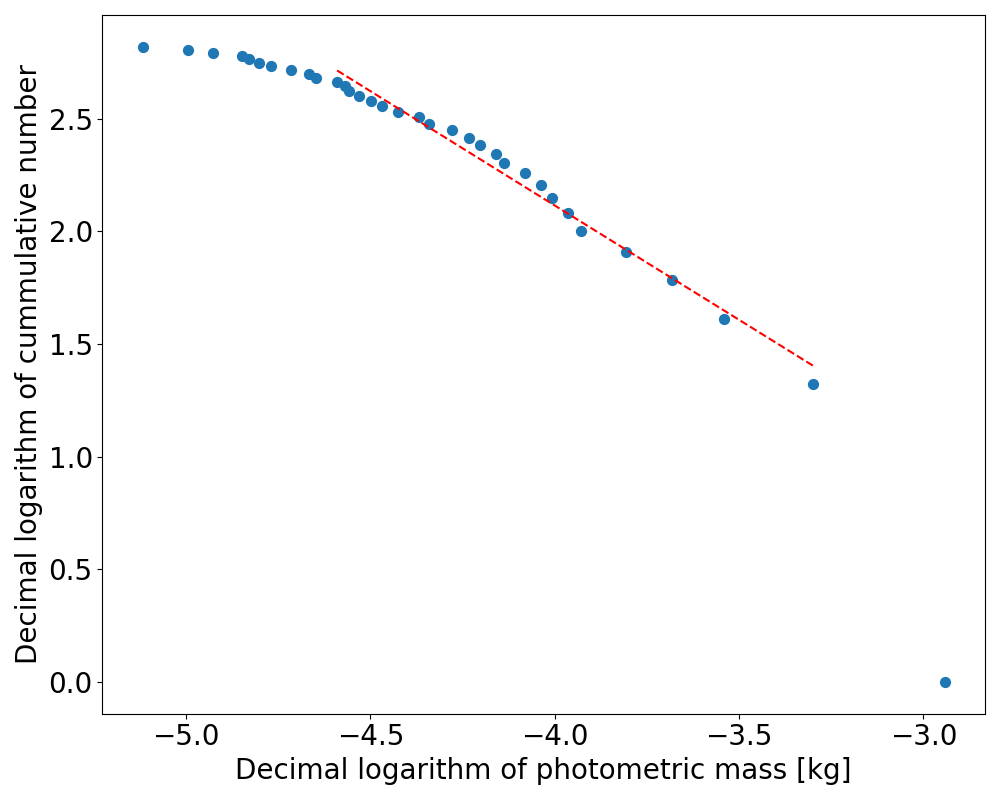}}
  \caption{Cumulative distribution of $\tau$~Herculid meteors recorded by the DMK and Maia cameras. The red line shows the fit of a linear part of the plot. The measured slope of the fit $k=-1.02$ is used for the mass distribution index calculation.} Each point represents the mean value for 20 meteors.
  \label{fig_mass_index_orig}
\end{figure}

The correction on different collection areas somewhat increased the number of meteors in the sample. Now, we work with an equivalent of 675 meteors, which were arranged according to their photometric mass. Mean masses of groups containing 20 meteors are plotted in Figure~\ref{fig_mass_index_orig}. The linear part of this plot was used for determination of the mass distribution index. The value $s = 2.02\pm0.12$ was found. The corresponding value of the population index is $r = 2.55$.

\begin{figure}
  \resizebox{\hsize}{!}{\includegraphics{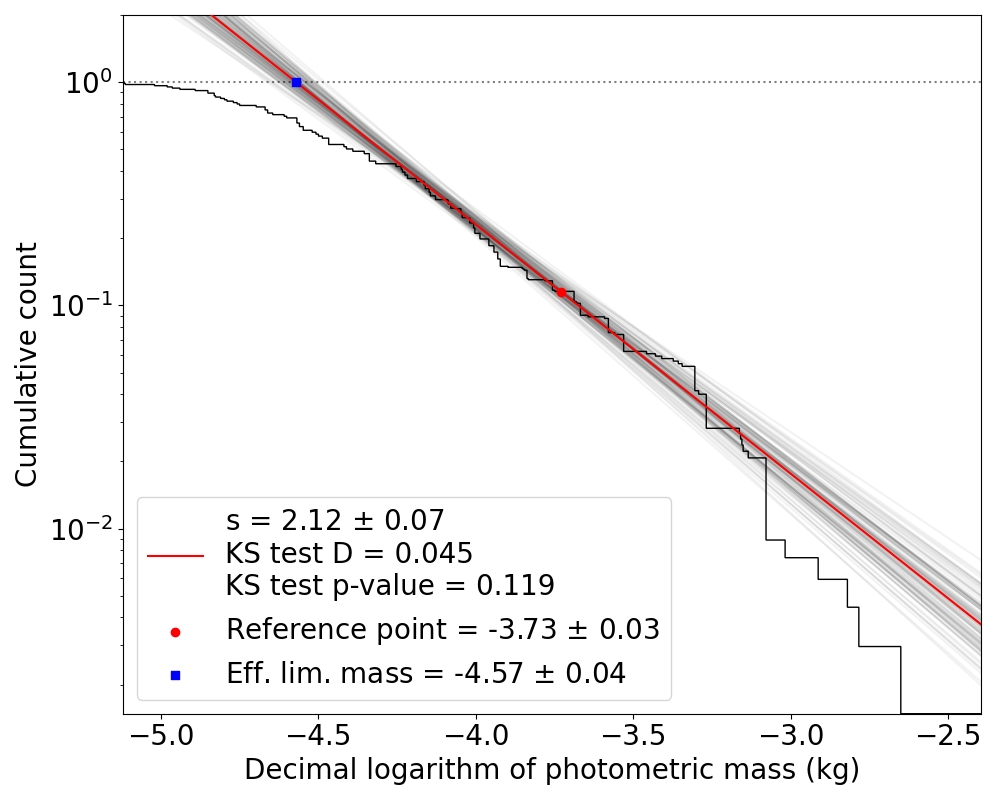}}
  \caption{The mass distribution index as computed by the software based on the MLE method \citep{Vida2020}.}
  \label{fig_mass_index_MLE}
\end{figure}

Recently, a robust method for the mass distribution index determination was developed by \citet{Vida2020}. It uses the maximum likelihood estimation (MLE) method to fit a gamma distribution to the observed distribution of masses. The source code is written in Python language and is available in Western Meteor Python Library\footnote{https://github.com/wmpg/WesternMeteorPyLib}. The {\tt FitPopulationAndMassIndex.py} software was applied on our data (corrected on different collection areas of cameras) and the resulting plot is seen in Figure~\ref{fig_mass_index_MLE}. This method provides a value $s = 2.12 \pm 0.07$.

The visual observation data results in $r = 2.40\pm0.06$ if the whole data set is used and $r = 2.57\pm0.23$ if only earlier activity is taken into account \citep{Rendtel2022}. Corresponding mass distribution indices are $s = 1.98$, respectively $s = 2.03$. The GMN preliminary report presents $r = 2.5\pm0.1$ with differential mass index $s = 2.0$ \citep{Vida2022}. Generally, it seems that all methods result in a mass distribution index slightly higher than 2.

\subsection{Meteoroid flux}
\label{flux}

To compute the flux of the meteoroids into the atmosphere it is necessary to know the collection area of each camera. It was calculated according to the method presented by \citet{Brown2002b}. This method neglects the Earth's curvature as well as certain decrease of camera sensitivity towards the edge of the field-of-view. A height of the maximum brightness of the meteors $H_{MAX}$ = 88~km was used for the collection area determination (see Section~\ref{heights}). The flux of the meteoroids producing meteors with brightness up to the camera meteor limiting magnitude $\Phi_{MLM}$ is computed as the corrected hourly rate (in 30 minute intervals) divided by this collection area. Finally, this value is transformed to the flux of the meteoroids up to +6.5 magnitude $\Phi_{6.5}$ using known values of the meteor limiting magnitude (MLM) of the camera and the population index (r) \citep{Brown2002}:

\begin{equation}
\Phi_{6.5} = \Phi_{MLM} . 10^{(6.5 - MLM) log r}
\end{equation}

Because the sky was cloudy at the Kun\v{z}ak observatory at the beginning of the experiment, the flux was determined only from the Ond\v{r}ejov video data. Values used for the calculation are summarised in Table~\ref{tab_flux_parameters}.

\begin{table}
\caption{Parameters used for the meteoroid flux calculation.}             
\label{tab_flux_parameters}      
\centering          
\begin{tabular}{l c c} 
\hline
camera						& 	Maia 			&	DMK 					\\
\hline
height of meteors			&	88 km			&	88 km					\\
camera elevation			&	45$^{\circ}$	&	45$^{\circ}$			\\
field-of-view diameter		&	52$^{\circ}$	&	22$^{\circ}$			\\
collection area				&	21500 km$^{2}$	&	2700 km$^{2}$			\\
meteor limiting magnitude	&	+5.5			&	+7.0					\\
population index			&	\multicolumn{2}{c}{2.55}					\\
\hline                  
\end{tabular}
\end{table}

\begin{figure}
  \resizebox{\hsize}{!}{\includegraphics{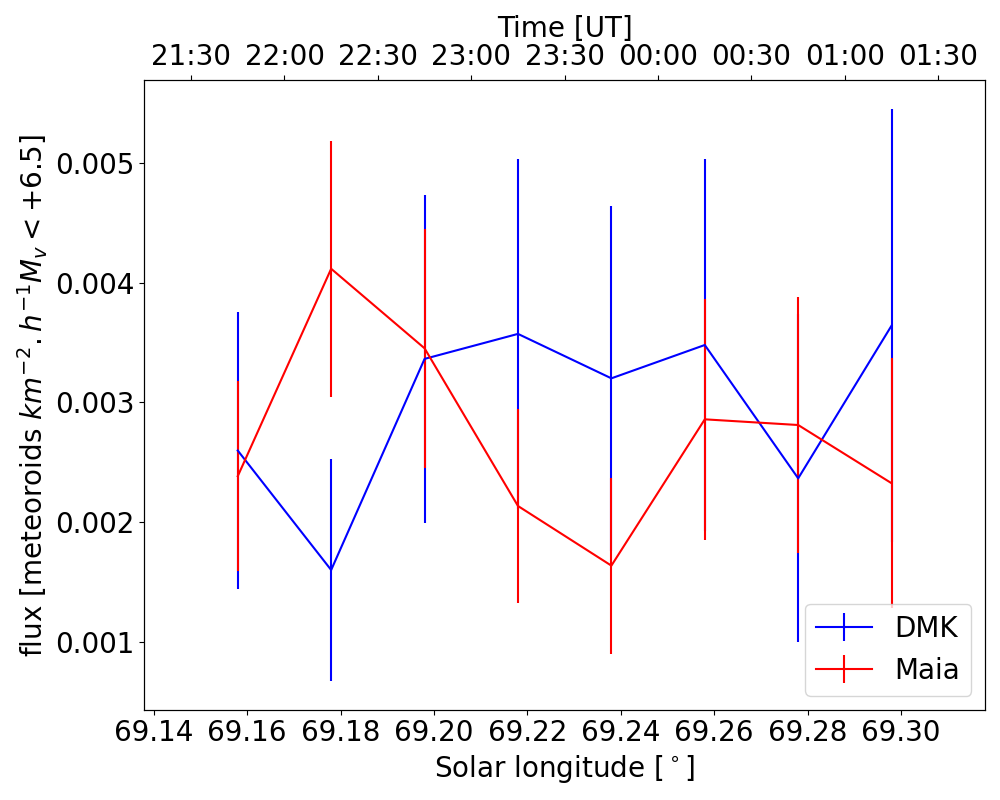}}
  \caption{A flux of $\tau$~Herculid meteoroids. Flux is calculated as a corrected hourly rate per km$^{2}$ per hour for meteoroids brighter than +6.5$^{mag}$. The blue line represents a flux as it was recorded by the DMK camera, whereas the red line as recorded by Maia camera, both located at Ond\v{r}ejov observatory.}
  \label{fig_flux}
\end{figure}

The flux of the meteoroids as recorded by both video cameras at Ond\v{r}ejov observatory is shown in Figure~\ref{fig_flux}. Although there are some fluctuations especially for the Maia camera, the values of the flux are relatively stable for several hours at around 0.0035$\pm$0.0013 meteoroids per km$^{2}$ per hour up to the +6.5 magnitude. The fluctuations are probably a consequence of the small number of the meteors. Less sensitive to the fluctuations seems to be a narrow field-of-view camera. The measured flux is approximately 3 times lower than the maximum flux reported by \citet{Vida2022} for the main peak of the shower activity. In comparison with other meteor shower outbursts, our measured value is about 10 times lower than the maximum flux of 2018 Draconids \citep{Koten2020}.

\section{Radiants and orbits}
\label{radiants}

There are not many $\tau$~Herculid meteor orbits published yet. Some photographic results were reported by \citet{Southworth1963} and \citet{Lindblad1971}. According to the IAU Meteor Data 
Center\footnote{\url{https://www.iaumeteordatacenter.org/}}, the radiant of the shower is $\alpha_{G}$=228$^{\circ}$.47, $\delta_{G}$=39$^{\circ}$.91. This is the radiant measured by \citet{Lindblad1971}. Hereafter we refer to these coordinates as catalogue coordinates of the radiant. Predictions for 2022 encounter differ from these values. They are summarised in Table~\ref{tab_radiants}.

\begin{table}
\caption{Summary of predicted and observed geocentric radiants for 2022 $\tau$~Herculid meteor shower.}             
\label{tab_radiants}      
\centering          
\begin{tabular}{l c c} 
\hline
							& $\alpha_{G}$ 		& $\delta_{G}$ 					\\
\citet{Luthen2001}			& 205$^{\circ}$.4	& 29$^{\circ}$.2			\\
\citet{Horii2008}			& 209$^{\circ}$.48	& 28$^{\circ}$.13			\\
\citet{Rao2021} 			& 210$^{\circ}$.17	& 25$^{\circ}$.03			\\
\citet{Ye2022}$^{a}$		& 209$^{\circ}$.4	& 28$^{\circ}$.3			\\
\hline
compact group (this work)	& 	209.17$^{\circ}\pm$0.14$^{\circ}$ 	&	27.86$^{\circ}\pm$0.12$^{\circ}$		\\
GMN \citep{Vida2022}		&	208.6$^{\circ}\pm$0.04$^{\circ}$	&	27.7$^{\circ}\pm$0.04$^{\circ}$		\\
CAMS \citep{Jenniskens2022}	&	209.17$^{\circ}\pm$0.09$^{\circ}$	&	28.21$^{\circ}\pm$0.07$^{\circ}$	\\
\hline
\end{tabular}
\tablefoot{$^{a}$ Model 2, assuming 2.5x Whipple's ejection velocity.}
\end{table}

\begin{figure}
  \resizebox{\hsize}{!}{\includegraphics{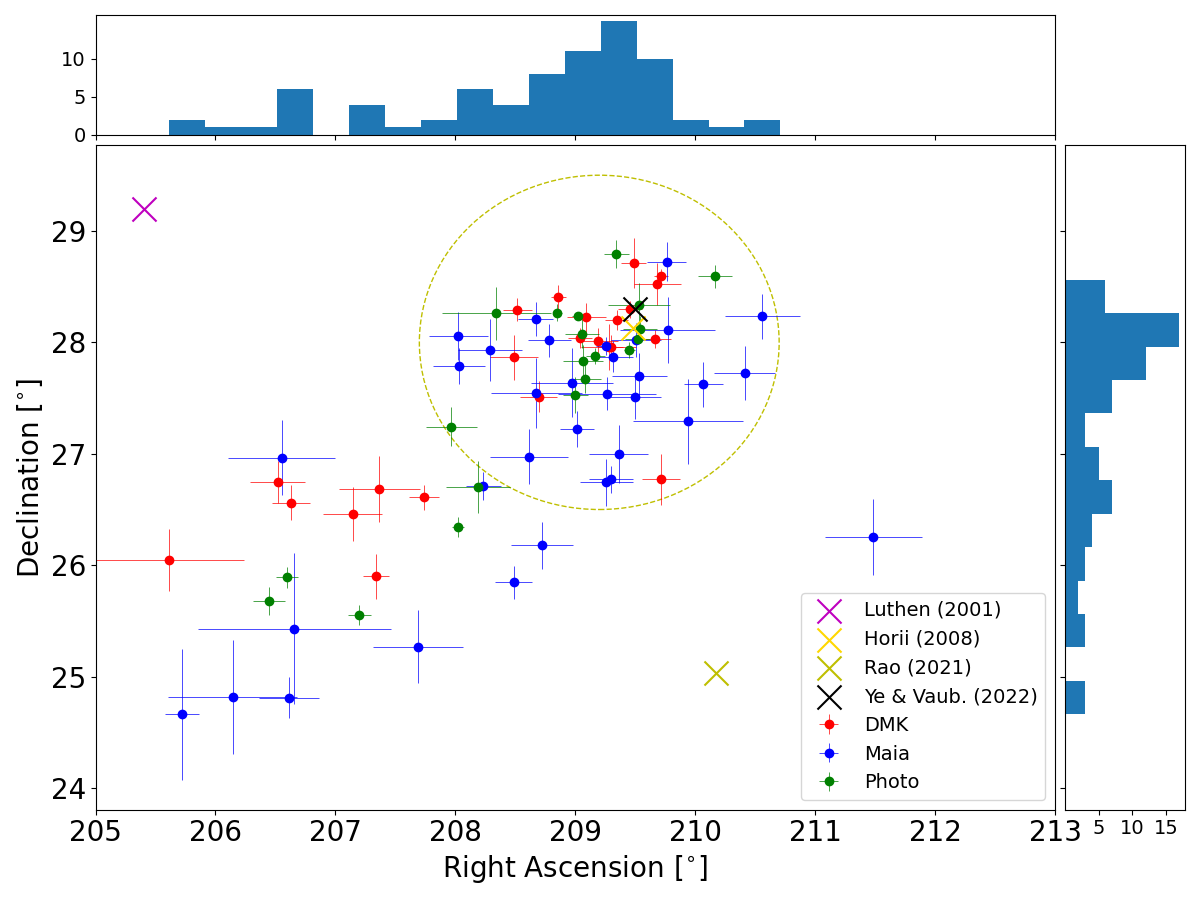}}
  \caption{The geocentric radiants of all observed $\tau$~Herculid meteors regardless of the technique used. Also radiant positions predicted by different models are included. Dashed yellow circle represents a compact radiant area with a radius of 1.5$^{\circ}$.}
  \label{fig_radiants}
\end{figure}

We obtained a double or multi station solution for 80 $\tau$~Herculid meteors. The geocentric radiants of all meteors together with the positions of the modelled radiants are shown in Figure~\ref{fig_radiants}. Two distinct areas can be seen here. Firstly, there is relatively compact group of radiants concentrated around  $\alpha_{G}$=209$^{\circ}$.2, $\delta_{G}$=27$^{\circ}$.9, secondly some radiants are scattered in south-western direction. The compact radiants are well consistent with the predictions of \citet{Horii2008} and \citet{Ye2022}.

The radiants are significantly shifted against the catalogue one. The angular distance of the centre of compact area from the catalogue value mentioned above is about 20$^{\circ}$. The distance of the scattered radiants is even higher. Actually, only two meteors with radiants relatively close to the catalogue one were recorded. These meteors are not included in our data sample, because they evidently do not belong to the analysed branch of the shower. As Figure~\ref{fig_radiants} also shows, the scattered radiants are not generally less precisely determined radiants. The error bars size is rather instrument dependent. The parameters of the photographic meteors are usually determined with lower errors in comparison with the video meteors.

\begin{figure}
  \resizebox{\hsize}{!}{\includegraphics{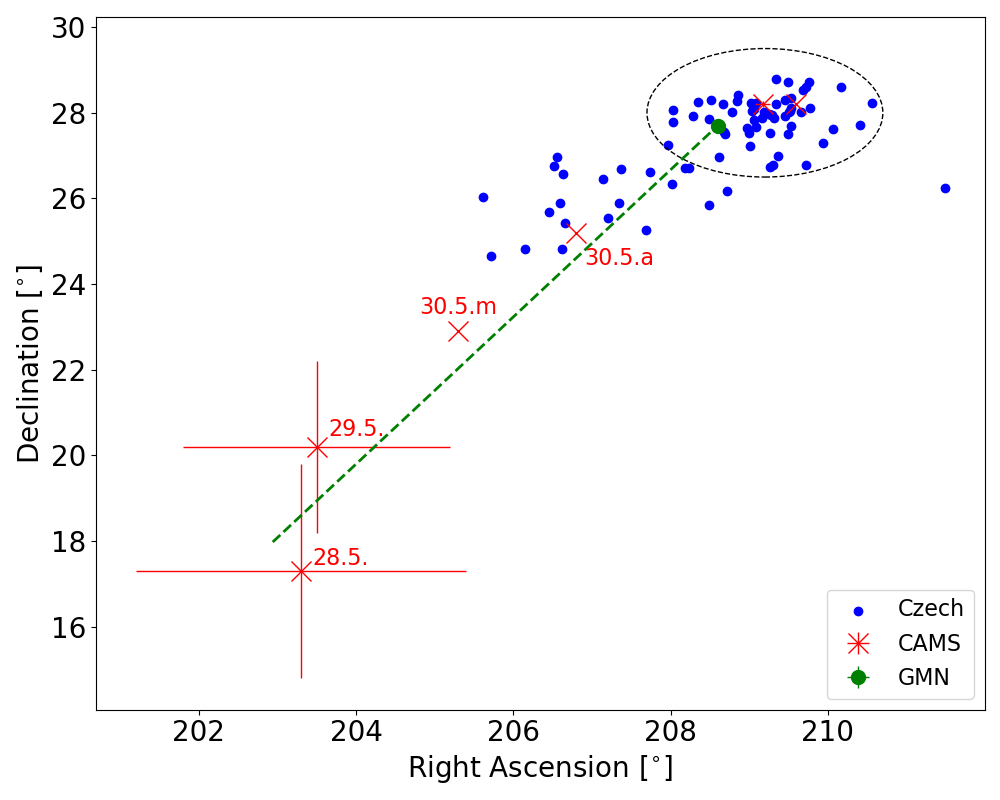}}
  \caption{Comparison of $\tau$~Herculid geocentric radiants observed by this experiment (blue circles), CAMS network (red crosses represents daily mean radiants) and GMN network (green circle represents mean peak radiant). The CAMS daily mean positions are additionally annotated by the date (m = morning, a = afternoon). Two positions without any annotation represent the evening and morning positions of the main peak night. Moreover, the radiant shift as recorded by the GMN is plotted for the last two days before the maximum (green dashed line).}
  \label{fig_radiants_comp}
\end{figure}

With known CAMS \citep{Jenniskens2022} and GMN \citep{Vida2022} networks data we can provide comparison of the results. The measured radiants by both experiments are also listed in Table~\ref{tab_radiants}. For CAMS data, the positions of the mean shower radiant are provided also for several days ahead of the maximum. On the other hand, for the GMN data the peak position is posted as well as the average radiant shift. As Figure~\ref{fig_radiants_comp} shows, the scattered radiants observed by our experiment are projected into the direction of the radiant shift. This could suggest that the meteoroid stream is wider and particles are more scattered within it.

\begin{table}
\caption{Heliocentric orbits of two groups of $\tau$~Herculid meteors and the parent comet.}             
\label{tab_orbits}      
\centering          
\begin{tabular}{l c c c} 
					&   compact          &    scattered 		&	73P \\
\hline
a [au]    			&   3.0$\pm$0.3    	 &    2.9$\pm$0.3		&	3.09	\\
e          			&   0.66$\pm$0.04	 &    0.65$\pm$0.03		&	0.686	\\
q [au]     			&   0.989$\pm$0.001  &    0.991$\pm$0.002	&	0.972	\\
i [$^{\circ}$] 		&	10.8$\pm$0.5     &    9.8$\pm$0.6		&	11.24	\\
$\omega[^{\circ}$]	&	200.0$\pm$0.4    &    199.6$\pm$0.8		&	199.39	\\
$\Omega[^{\circ}$]  &	69.24$\pm$0.05   &    69.24$\pm$0.06	&	69.66	\\
\hline
\end{tabular}
\end{table}

An explanation for two groups of the radiants provides a paper of \citet{Egal2023}. The compact area of the radiants is well consistent with the modelled radiants originated from the 1995 comet break-up. On the other hand, the scattered radiants agree with the simulated meteoroids, which were released between 1900 and 1947.

\begin{figure}
  \resizebox{\hsize}{!}{\includegraphics{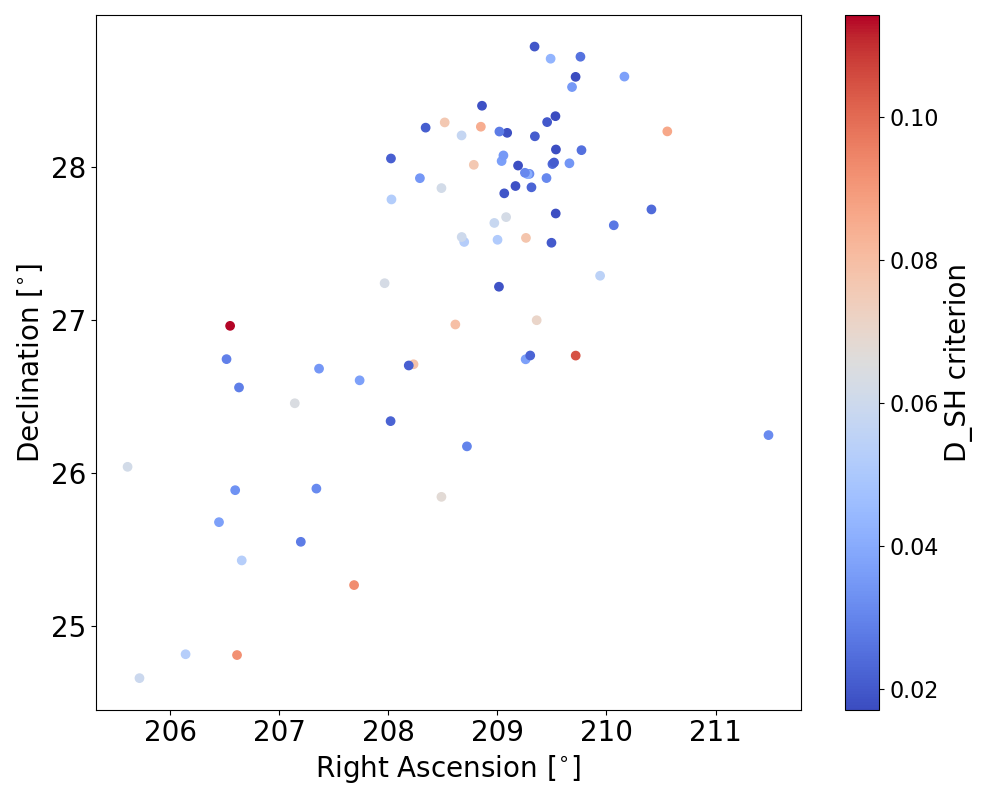}}
  \caption{The distribution of the geocentric radiants of $\tau$~Herculid meteors with colour-coded $D_{SH}$ criterion against the parent comet orbit.}
  \label{fig_radiants_DSH}
\end{figure}

Several quantities were tested to see whether they influence the position of the radiants in the plot. The photometric mass of the meteoroids, its semimajor axis, perihelion distance, time of its appearance expressed in terms of the solar longitude, the inclination of the orbit as well as $D_{SH}$ criterion were among the tested parameters. As an example a distribution based on the $D_{SH}$ criterion against the 73P/Schwassmann-Wachmann~3 orbit is used in Figure~\ref{fig_radiants_DSH}. It is clearly visible that the position of the radiant does not depend on the distance between comet and meteoroid orbits. Moreover, this figure also shows that the heliocentric orbits of the majority of meteors are very close to the parent comet orbit with $D_{SH}$ < 0.06 (shades of blue). 

\begin{figure}
  \resizebox{\hsize}{!}{\includegraphics{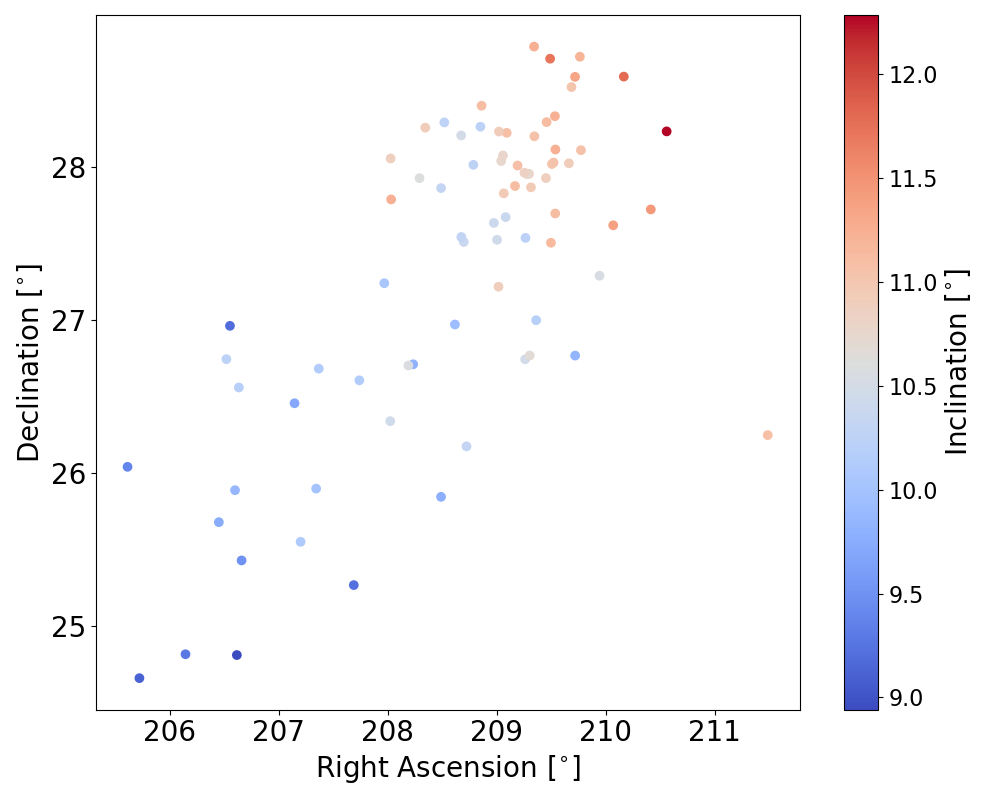}}
  \caption{The distribution of the geocentric radiants of $\tau$~Herculid meteors with colour-coded inclination of the heliocentric orbit.}
  \label{fig_radiants_incl}
\end{figure}

No dependency on the majority of tested parameters was found. The only exception seems to be the inclination of the heliocentric orbit as Figure~\ref{fig_radiants_incl} suggests. There are meteors with lower value of the inclination among the scattered radiants, whereas the compact radiant area is dominated by the meteors with inclination higher than 10.5$^{\circ}$.

\begin{figure}
  \resizebox{\hsize}{!}{\includegraphics{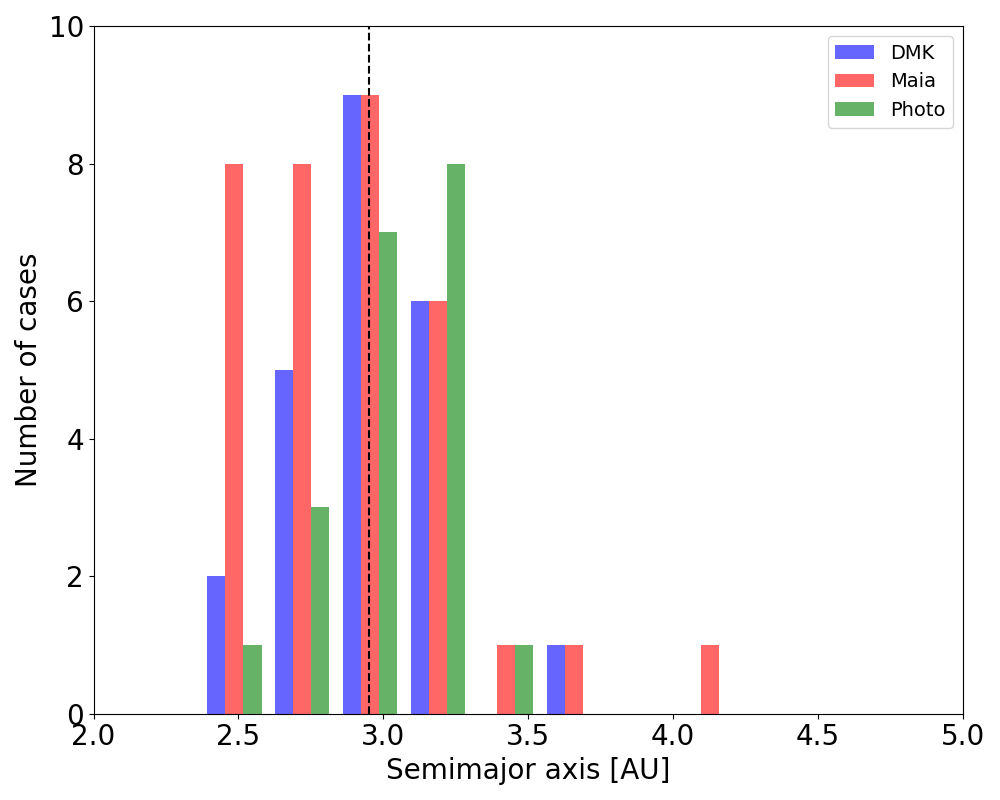}}
  \caption{The histogram of the semimajor axis distribution of $\tau$~Herculid meteors observed by different kinds of the instruments with mean value marked by the dashed black line.}
  \label{fig_axis_hist}
\end{figure}

We compared the heliocentric orbits of both groups with the orbit of the parent comet. The results are shown in Table~\ref{tab_orbits}. With the exception of the argument of perihelion, the values of the compact group are a bit closer to the comet values. Higher inclination of the compact radiants is also seen from this table. Mean semimajor axis is slightly smaller in comparison with the parent comet for both groups of the meteors. The histogram of the semimajor axis distribution is seen in Figure~\ref{fig_axis_hist}. A dependance of the semimajor axis on the photometric mass was not found.

\section{Physical properties}
\label{properties}

Our sample of $\tau$ Herculid meteors covers a range of the maximum brightness between 4.8 and -11.4 magnitudes. The corresponding photometric masses range from 7.7~mg to about 10~kg, although the upper limit is probably even higher. This fact allows us to investigate properties of a broad range of meteoroid masses which reach almost 7 orders of magnitude. 

\subsection{Height data}
\label{heights}

The height data can be used as an insight into the meteoroid structure. Generally, three different kinds of instruments were used for the observation, each of them is characterised by different sensitivity. It means that the beginning heights of e.g. photographic meteors cannot be directly compared with the video meteors. Also both video experiments differ in their sensitivity. Therefore, the double station meteors detected by Maia cameras were selected for this analysis. They provide a slightly larger sample of uniform data than DMK or photographic cameras. 

\begin{figure}
  \resizebox{\hsize}{!}{\includegraphics{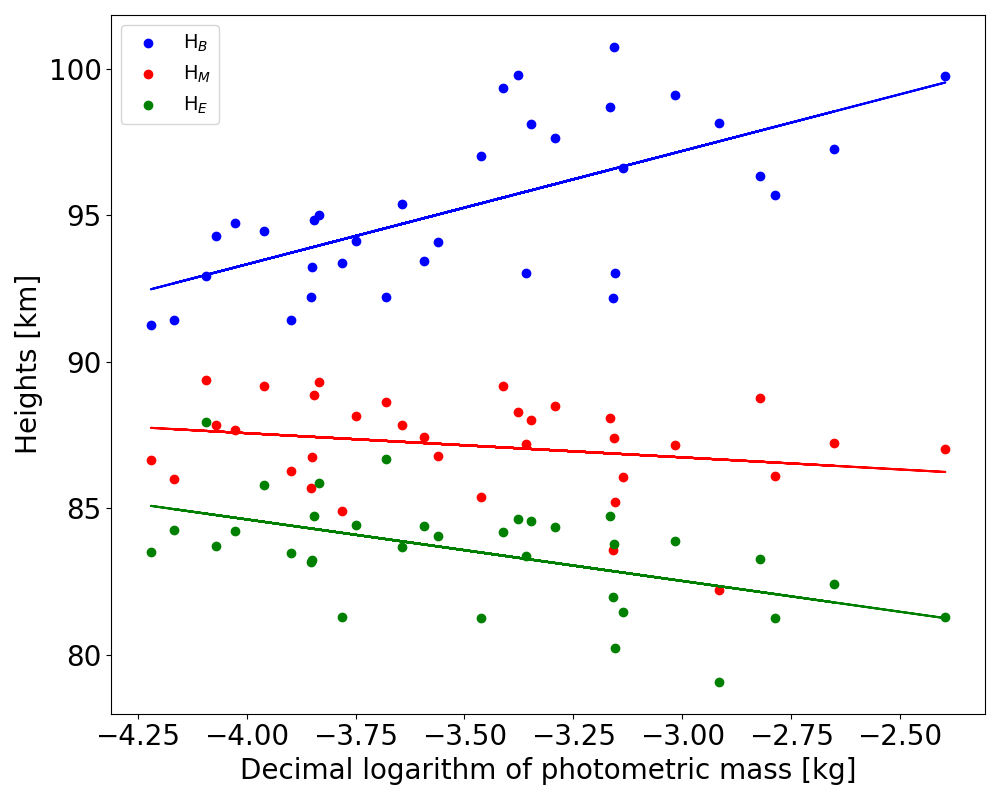}}
  \caption{The beginning height, height of maximum brightness and terminal height of $\tau$~Herculid meteors detected by the Maia cameras.}
  \label{fig_heights}
\end{figure}

The beginning height, height of the maximum brightness and the terminal height are plotted as a function of the photometric mass in Figure~\ref{fig_heights}. The beginning heights are generally between 100 and 90 km. There is an increase in the beginning height with increasing photometric mass. The slope of the linear fit is about 3.9. This result is in agreement with the conclusion of \citet{Koten2004} who found that the beginning heights of cometary meteors increase with increasing photometric mass. On the other hand, when comparing the slope of the fit with values presented in the mentioned paper for such showers as Perseids, Taurids and Orionids, the value found for $\tau$ Herculids is a bit smaller. The dependence of the beginning height on the photometric mass is shallower in comparison with other cometary meteor showers. On the other hand, it is still significantly higher than in the case of Geminid meteors, for which a completely different composition was found.

The height of maximum brightness as well as the end height both decrease with increasing photometric mass. The linear fit for the latter one is a bit steeper. It means that heavier meteoroids penetrate deeper into the atmosphere. The maximum brightness is usually between 90 and 85~km. The average value is $H_{MAX}$ = 88~km (value used for the collection area calculation). The deepest penetration is about 80~km. The heights at which the $\tau$ Herculid meteors occurred are higher than expected for such slow meteors. Note, that when the observation campaign was planned, the aiming point for the double station video observation was set at 80~km. 

\subsection{$K_{B}$ parameter, PE criterion}
\label{KB-PE}

The altitude data are influenced by the slope of the meteor trajectory as well as the initial speed. Therefore, several dimensionless criteria were defined to compensate for these effects. Summary of them can be found in \citet{Cep1988}. For faint television and video meteors, the $K_{B}$ parameter is used, which corrects the beginning height for the initial speed and slope of the trajectory. The meteoroids are classified into groups A, B, C, and D depending on $K_{B}$ value. Subgroups C1, C2, and C3 are defined according to the meteoroid orbit. The $PE$ and $AL$ criteria are used for fireballs. Here, we apply the PE criterion, which compensates the terminal height for the meteoroid mass, initial speed and slope of the trajectory. Classification scheme consisting of groups I, II, IIIA, IIIB was introduced according to this criterion. 

\begin{figure}
  \resizebox{\hsize}{!}{\includegraphics{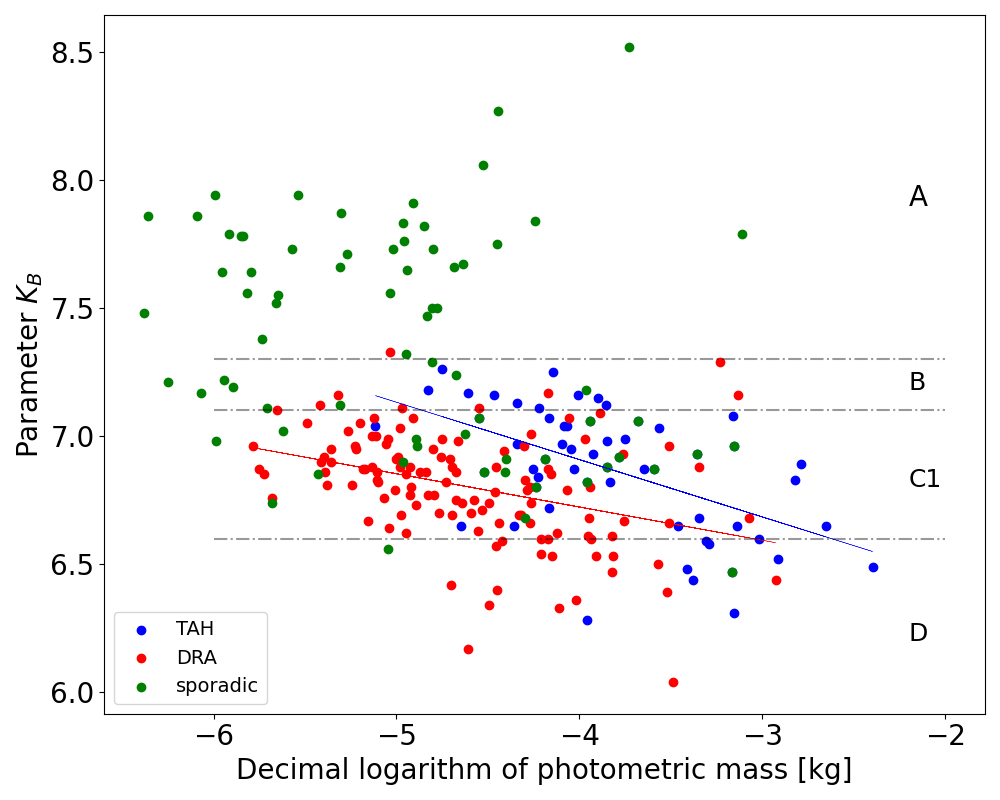}}
  \caption{The $K_{B}$ parameter of $\tau$ Herculid, Draconid and slow sporadic meteors ($v_{\infty} < 20$ km/s) recorded by the same kind of video cameras with limits for the classification scheme. Trends of $K_{B}$ for $\tau$ Herculids and Draconids are also shown.}
  \label{fig_KB}
\end{figure}

Figure~\ref{fig_KB} shows distribution of $K_{B}$ parameter for $\tau$ Herculid meteors. The values range between 6.3 and 7.3 with a mean value $\overline{K_{B}} = 6.9 \pm 0.2$. There are only 10 meteors belonging to group B which represents dense cometary material. It is a bit surprising that such material is present within this cometary meteor shower. Therefore, records of these cases were checked again and it was confirmed that real beginning is measured for all of them. Majority of the meteors fall into C group, i.e. regular cometary material. With a given semi-major axis and inclination, C1 subgroup is occupied by $\tau$ Herculid meteors. Finally, a small portion of the meteors belongs to group D, representing soft cometary material.

We can compare the $\tau$ Herculid meteors with other meteor showers or sporadic meteors. Included in the Figure~\ref{fig_KB} are also $K_{B}$ parameters of 2018 Draconids and of slow sporadics ($v_{\infty} < 20$ km/s). Both sets of data were obtained by the same video system. The Draconids\footnote{009 DRA code in IAU MDC database} are usually supposed to be very fragile meteoroids. On the other hand, their speed ($v_{\infty} \sim 23.6$ km/s) is higher than in the case of $\tau$ Herculids ($v_{\infty} \sim 16$ km/s). For smaller masses down to 10~milligrams their $K_{B}$ parameters are similar to $\tau$ Herculid meteors. Generally, both shower meteoroids are softer than the sporadic meteors with similar velocity as the many sporadic meteors are of asteroidal origin.

Although the values of $K_{B}$ criterion are scattered in a broad range for both meteor showers, there is a trend to lower values of $K_{B}$ criterion with increasing photometric mass. This trend is steeper for $\tau$ Herculid meteors. Among smaller masses, Draconid meteoroids are softer but as we approach higher masses, the $K_{B}$ are getting closer and for the highest masses in the sample they become very similar.

\begin{figure}
  \resizebox{\hsize}{!}{\includegraphics{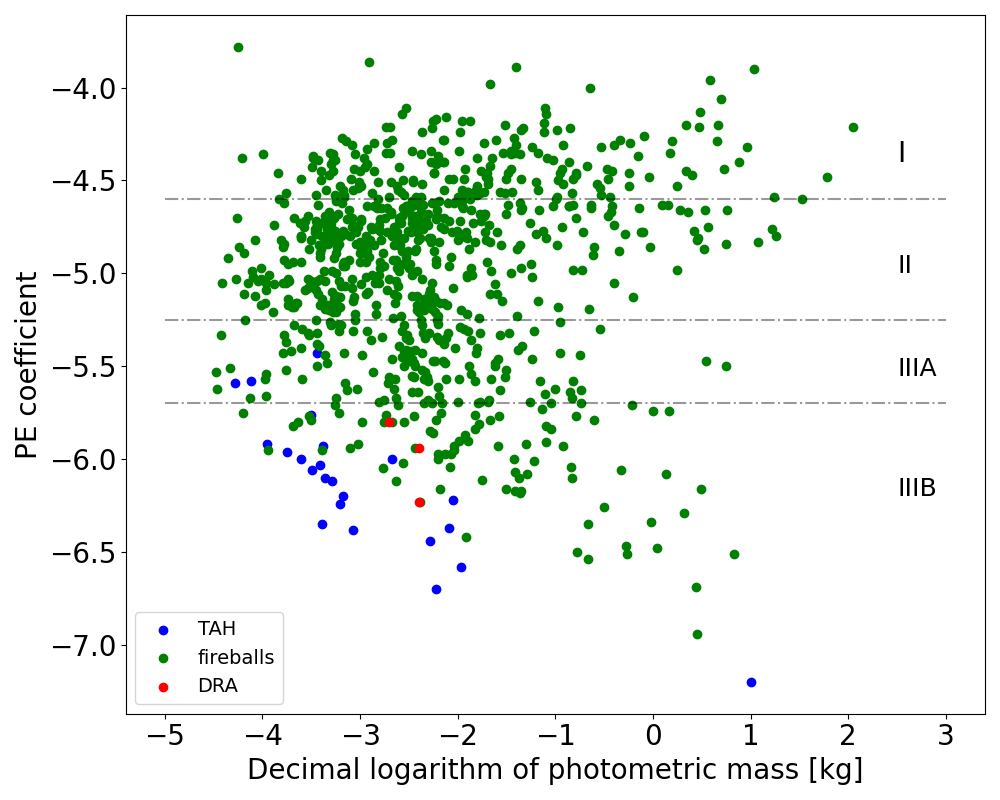}}
  \caption{The PE criterion of $\tau$ Herculid photographic meteors (blue) and their comparison with Draconid (red) and other fireballs (green) from the catalogue of \citet{Borovicka2022b}.}
  \label{fig_PE_comp}
\end{figure}

Classification of $\tau$ Herculid photographic meteors according to PE criterion is shown in Figure~\ref{fig_PE_comp} where different groups are also marked. We can see that almost all meteors belong to IIIB (soft cometary material), only few of them belong to groups IIIA (regular cometary material). There is also a very clear trend in PE values depending on the mass of meteoroids, i.e. more massive meteoroids are more fragile. The PE criterion for 3 Draconid and other fireballs observed in 2017--2018 \citep{Borovicka2022b} shows that our observed $\tau$ Herculids are the most fragile meteoroids observed ever. Their PE is usually lower in comparison with other meteors and even Draconid meteors of the same mass.

\subsection{EN310522\_013924 - exceptional case of bright $\tau$~Herculid bolide}
\label{EN310522_013924}

As is already seen from Figure~\ref{fig_PE_comp} there is a very bright and fragile bolide among the photographic meteors. This one was recorded by four all-sky photographic cameras and four IP video cameras at a total of 6 stations of the European fireball network (EN) in the Czech Republic (3), Slovakia (2) and Austria (1) on 31 May 2023 at 1:39:22.5 $\pm$ 0.5~UT (for bolide beginning). At that time, it was already late dawn especially at the easternmost stations. As the bolide was flying over southeastern Europe (Bosnia and Herzegovina), it was only detected at the southern EN stations and everywhere it was very low above the horizon. Nevertheless, thanks to the high quality and high-resolution recordings, all the parameters of its passage through the atmosphere could be determined with very good accuracy and reliability. The closest station (with a mean distance from the bolide of 465 km) from where we have both an all-sky photographic image and a video recording was Hurbanovo in southern Slovakia. The farthest station (mean distance 755 km) was Ond\v{r}ejov (IP camera). The fact that the bolide was exceptionally bright is proved by the fact that it was well recorded at such a large distance and also during dawn. This is confirmed by the analysis of the above records from which we determined that the bolide reached maximum absolute brightness of -11.6~magnitude and the initial meteoroid mass was 10~kg or slightly more. It should be emphasised here that the brightness and mass of the bolide were the most difficult of all parameters to determine, mainly because of the already very over-illuminated sky, and thus may be affected by a somewhat larger error than all other parameters. It is also likely that the reported values of brightness and mass are somewhat underestimated rather than unrealistically high.

The analysis of physical properties of this meteoroid clearly shows that it was an extremely fragile material. If we take into account the input mass mentioned above and the atmospheric trajectory parameters presented in Table~\ref{tab_bolide_traj}, the PE coefficient \citep{Cep1988} describing the material properties of the body comes out to -7.31, which is such a low value that we have not observed for any other bolide in the entire existence of the European fireball network. This indicates not only the extreme fragility of this meteoroid, but also its extremely low bulk density. For the most fragile meteoroids with PE less than -5.70, a bulk density of 270~kg.m$^{-3}$ \citep{Cep1988} is usually given, but in this case, it must have been even much lower. \citet{Egal2023} modelled light curves of two meteoroids detected by CAMO and found bulk densities even lower -- down to 230~kg.m$^{-3}$. However, even if we take a density of 270~kg.m$^{-3}$, the diameter of this meteoroid would be about 42~cm at the given mass. Obviously, this is only a lower estimate and the real size of this meteoroid could easily have been around half a metre or even larger. 

\begin{table*}
\caption{Atmospheric trajectory of EN310522\_013924 bolide.}             
\label{tab_bolide_traj}      
\centering          
\begin{tabular}{l | c c } 
\hline
Atmospheric trajectory			&	Beginning				&	Terminal				\\
\hline
Height [km]						& 	99.32 $\pm$ 0.09		&	76.72 $\pm$ 0.07		\\
Longitude [$^{\circ}$ E]		&	18.6606 $\pm$ 0.0012	&	19.0591 $\pm$ 0.0010	\\
Latitude [$^{\circ}$  N]		&	43.8694 $\pm$ 0.0010	&	43.7769 $\pm$ 0.0008	\\
Slope [$^{\circ}$]				&	33.6 $\pm$ 0.3			&	33.4 $\pm$ 0.3			\\
Azimuth [$^{\circ}$]			&	107.6 $\pm$ 0.3			&	107.9 $\pm$ 0.3			\\
Total length [km]/Duration [s]	&	\multicolumn{2}{c}{40.93/2.55}						\\
EN stations and cameras			&	\multicolumn{2}{l}{Hurbanovo (all-sky+IP), Kun\v{z}ak (all-sky+IP),} 		\\
								&	\multicolumn{2}{l}{Rimavsk\'{a} Sobota (all-sky), Martinsberg (all-sky),}	\\
								& 	\multicolumn{2}{l}{Vesel\'{i} nad Moravou (IP), Ond\v{r}ejov (IP)}			\\
\hline
\end{tabular}
\end{table*}

As explained above, unlike the brightness and mass, the trajectory is determined to a very good accuracy with a standard deviation of 65~metres for any point along the luminous path, and the initial velocity even with a standard deviation of only 30~m s$^{-1}$ (Table~\ref{tab_bolide_orbit}). A detailed analysis shows that the meteoroid entered the atmosphere at 16.0~km s$^{-1}$ and started its light, i.e. began to be visible for our cameras at an altitude of 99.3~km. It travelled a 40.93~km long luminous trajectory in 2.55~s and terminated its flight at an altitude of 76.7~km. The entire atmospheric trajectory was over southeastern Bosnia and Herzegovina (east of Sarajevo). Data describing the atmospheric trajectory as well as its heliocentric orbit are given together with the corresponding standard deviations in Table~\ref{tab_bolide_orbit}. Similar to the atmospheric trajectory, the heliocentric orbit of this surprisingly large $\tau$~Herculid is determined reliably and matches very well the orbit of comet 73P/Schwassmann-Wachmann~3. The similarity criterion $D_{SH}$ is only 0.023 (cut-off value for $D_{SH}$ is = 0.20), and the other similarity criterion $D_{D}$ defined by \citet{Drummond1981} is only 0.012 (cut-off  value for $D_{D}$ is = 0.105). Both values demonstrate a clear orbital connection between this meteoroid and the parent comet 73P/Schwassmann-Wachmann~3. Its radiant lies in the compact area.

\begin{table}
\caption{Radiants and orbital elements of EN310522\_013924 bolide (J2000.0).}             
\label{tab_bolide_orbit}      
\centering          
\begin{tabular}{l | c} 
\hline
$\alpha_{R}$ [$^{\circ}$]	& 	217.6 $\pm$ 0.4			\\
$\delta_{R}$ [$^{\circ}$]	& 	34.55 $\pm$ 0.18		\\
$v_{\infty}$ [km s$^{-1}$]	&	15.97 $\pm$ 0.03		\\
$\alpha_{G}$ [$^{\circ}$]	& 	208.1 $\pm$ 0.4			\\
$\delta_{G}$ [$^{\circ}$]	& 	28.2 $\pm$ 0.2   		\\
$v_{G}$ [km s$^{-1}$]		&	11.86 $\pm$ 0.04		\\
$\lambda_{E}$ [$^{\circ}$]	& 	167.15 $\pm$ 0.08		\\
$\phi_{E}$ [$^{\circ}$]		& 	10.69 $\pm$ 0.09		\\
$v_{E}$ [km s$^{-1}$]		&	38.23 $\pm$ 0.05		\\
a [au]						&	3.07 $\pm$ 0.04			\\
e							&	0.677 $\pm$ 0.004		\\
q [au]						&	0.9913 $\pm$ 0.0005		\\
Q [au]						&	5.15 $\pm$ 0.09			\\
$\omega$ [$^{\circ}$]		&	199.1  $\pm$ 0.2		\\
$\Omega$ [$^{\circ}$]		&	69.3295 $\pm$ 0.0001	\\
i [$^{\circ}$]				&	10.91 $\pm$ 0.09 		\\
P [y]						&	5.38 $\pm$ 0.11			\\
T$_{J}$						&	2.81 $\pm$ 0.02			\\
\hline
\end{tabular}
\end{table}

The similarity of the orbit of such a massive and large meteoroid to the parent comet poses a significant challenge for modelling the evolution of this meteor shower. Majority of models mentioned above predicted that the particles released in 1995, which could reach the Earth's orbit in 2022, could only be up to a millimetre in size. \citet{Egal2023} explained the observation of several bright fireballs using particles of a few centimetres which were released from the comet during the 1995 fragmentation event or during previous apparitions, but a half metre size body still needs to be simulated. From this point of view, this is a very important observation that brings new and unexpected insights into the description of the dynamics of this meteor shower.

\subsection{Meteor spectra}
\label{spectra}

Meteor spectra can be used to study the composition of meteoroids and also their physical structure. Preferential release of sodium is a sign of early fragmentation into small grains \citep{Borovicka2014}. The spectral video camera captured only two spectra of sufficient quality. The spectrum of meteor 22530019 (hereafter m019) was observed on the non-blazed side of the grating  ($-$1st order), which provides a much lower signal in the blue and green part of the spectrum. The spectrum of meteor 22530038 (hereafter m038) was observed in the +1st order but only until the wavelength of 610~nm. In addition, the spectrum of the terminal flare (magnitude $-7.7$) of the fireball EN300522\_214616 (hereafter f214616) was obtained by a SDAFO. There were some other spectra obtained but they show only the Na line. It is valid also for the spectrum of the very bright and distant fireball EN310522\_013924 (see \ref{EN310522_013924}), which was captured by the IP camera in Kun\v{z}ak.

\begin{figure}
  \resizebox{\hsize}{!}{\includegraphics{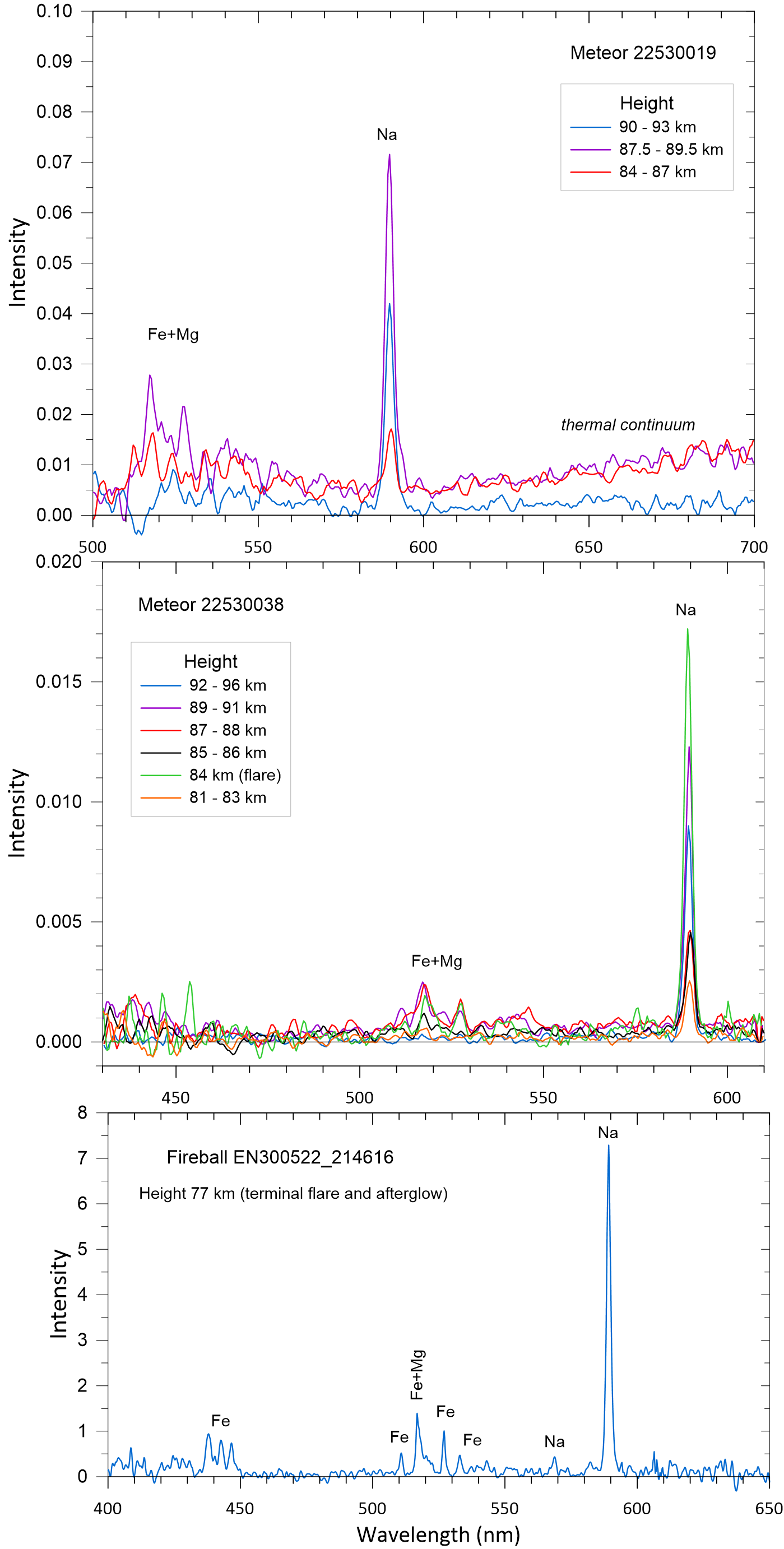}}
  \caption{Calibrated spectra of meteors 22530019 and 22530038 at various height ranges and the time-integrated spectrum of the fireball EN300522\_214616 at the position of its terminal flare. The intensities are given in kW nm$^{-1}$ ster$^{-1}$.}
  \label{fig_spectra}
\end{figure}

The spectra of both video meteors evolved in time. At the beginning, only the Na line (589 nm) was present. Only when it became quite bright, the lines of Fe and Mg at 510--550 nm and a continuum appeared. The Na line then started to fade, while the Fe and Mg lines and the continuum kept their brightness. The continuum was especially well visible in m019 above 600~nm. The continuum showed no band structure, so it was a thermal continuum (with Planck temperature of about 2000~K) and not a molecular radiation such as N$_2$ or FeO observed in other meteors. While the Na line disappeared shortly before meteor ended in m019, in m038 it retained some residual brightness. The other lines faded and Na became the dominant spectral feature again, though at lower intensity than at the beginning. Then, at the height of 84 km, a flare occurred. The Na line was very bright in the flare and remained visible also after the flare at lower heights. The spectra at different heights are plotted in Fig.~\ref{fig_spectra}. Note that the spectrum of the flare is more noisy (especially around 450 nm), since it is a single frame spectrum while other spectra have been averaged over several frames.

The photographic spectrum of f214616 was taken at a single height. Since the exposure was long, it contains both the spectrum of the fireball during the flare and the spectrum of a possible afterglow, which likely remained visible for some time at the same spot. Afterglows are typical by the presence of low excitation Fe lines (multiplets 1 and 2) and they are clearly seen in the spectrum. This spectrum has higher resolution than the video spectra. It was possible to confirm that Fe lines are brighter than Mg lines at 516--519 nm. But the brightest line was again the Na line.

\citet{Borovicka2014} have shown, on the example of the Draconids, that meteoroids of the same stream observed in a single night can have different structure. Some meteoroids disintegrated into grains at the beginning of the trajectory, some were more resistant. Since sodium easily evaporates from small grains, it appears and disappears in the spectrum earlier than other elements in the former case. Meteoroid m019 was such a case. Meteoroid m038 was a combined case. A larger part disintegrated at the beginning, at height around 96 km, another part remained compact until the height 84 km, where it erupted, and a small part continued even lower. 

When integrating line intensities along the whole trajectory, m038 can be classified as enhanced Na in the scheme of \citet{Borovicka2005}. In fact, in all $\tau$ Herculids with spectra, the Na line was dominating and the Mg line was much fainter than in Draconids for example. This fact can be partly connected with the lower velocity of $\tau$ Herculids. But it is possible that they were indeed richer in volatile sodium because they spent only 27 years in the interplanetary space and were buried inside the comet before that.

\section{Discussion}
\label{discussion}

Number of models predicted possible activity of the $\tau$ Herculid meteor shower in morning hours of 31 May, 2022 caused by the material released from the parent comet during its 1995 outburst (for summary see Table~1 in \citet{Ye2022}). Earlier activity connected with this disruption event was not predicted. There was only one older model by \citet{Wiegert2005}, which expected meteor shower activity before midnight, i.e. in the very evening of 30 May. This activity should be connected with material released from the comet in 1892 and 1897.

Our video and photographic observations as well as the visual data of the IMO observers \citep{Rendtel2022} and video data of the GMN \citep{Vida2022} registered higher activity well above the annual level during the last hours of 30 May. Taking into account model predictions, this activity may be connected with 19th century perihelion passages of the parent comet. In terms of activity level, the event was smaller in comparison with other recent events such as the modest outburst of the 2018 Draconid meteor shower. On the other hand, the activity was significant for the usually almost missing $\tau$ Herculid meteor shower and provided us with a lot of valuable data.

An important contribution to determination of the observed meteors origin was provided by the models of \citet{Egal2023}. They simulated particles released from the parent comet at each apparition from 1800 as well as during the 1995 fragmentation of the comet. The modelled activity profile shows a secondary maximum between 15 and 19~UT on May 30 as well as the main peak after 4~UT on May 31. According to those models, our observation window fits between both peaks. The earlier peak was at its descending branch, whereas there was already ascending activity connected with the main peak. The model also shows that the main peak was wider than originally expected. Therefore, we were able to record also meteors belonging to the 1995 ejecta although the observation period started hours before the predicted maximum.

\begin{figure}
  \resizebox{\hsize}{!}{\includegraphics{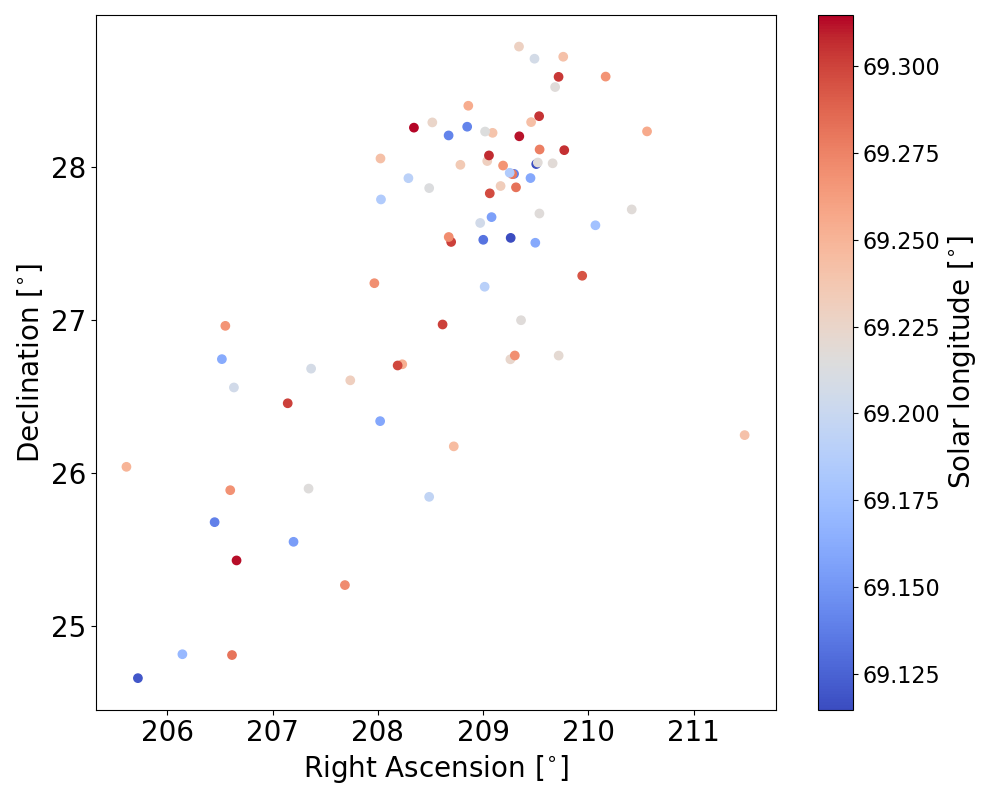}}
  \caption{The distribution of the geocentric radiants of $\tau$~Herculid meteors with colour-coded solar longitude. The scattered and compact radiants are distributed along a whole observation period.}
  \label{fig_radiants_sollong}
\end{figure}

The fact that we observed a mix of older meteoroids released before 1947 and young particles from the 1995 fragmentation is supported also by comparison of the recorded and simulated radiants. Figure~\ref{fig_radiants_sollong} shows that meteors recorded along a whole observation period contributed to both groups of radiants. The models of \citet{Egal2023} provide very good agreement with our results. The scattered radiants are consistent with older meteoroids, the compact radiants with fresh ones. Also the heliocentric orbits of the majority of observed meteors are very close to the current orbit of the parent comet. 

The analysis of the physical properties shows that $\tau$ Herculid meteoroids are generally very fragile. Moreover, we can say that their fragility increases with increasing mass of the meteoroid. When looking at the smallest masses, the particles seem to be a bit stronger than Draconid meteoroids. With an increasing mass both groups of meteoroids become more similar. The trend of increasing fragility continues into the bigger masses. $\tau$ Herculid meteors become, according to their PE criteria, even more fragile than the Draconids. When compared with other meteoroids of the same mass their PE values are the lowest recorded ever.

The recorded spectra confirmed early disruption of meteoroids into grains but also showed the existence of more resistant parts. A detailed meteoroid fragmentation modelling is planned for the future. The spectra also suggest an enhanced abundance of volatile sodium, which may be related to the short time the meteoroids were exposed to solar and cosmic radiation.

There are significant implications for the theoretical models of meteoroid stream formation and evolution. Much heavier particles than the models took into account were presented. Initially, according to models, only millimetre size meteoroids were able to reach an Earth's orbit. Already \citet{Ye2022} note that the observed meteoroids were heavier and suppose that the brighter meteors could be caused by centimetre size, porous dust aggregates. \citet{Egal2023} also took into account particles of a few centimetres. Still, much bigger than centimetre size particles were observed. The 10~kg meteoroid was almost a half metre size body. This big body surely belongs to the 1995 fragmentation event. It is necessary to explain how such meteoroids could get close to the Earth at the same time  as the millimetre size particles did.

\begin{acknowledgements}
This work was supported by the Grant Agency of the Czech Republic grants 20-10907S (video observation and analysis), 19-26232X (photographic observation and analysis) and the institutional project RVO:67985815 (institutional infrastructure). We thank the members of the Czech Hydrometeorological Institute for the forecast service in the days before the observations.
\end{acknowledgements}

\bibliographystyle{aa}
\bibliography{00pkbib}

\begin{appendix}

\section{Geocentric radiants and heliocentric orbits of $\tau$-Herculid meteors}

\begin{table*}
\caption{Geocentric radiants of $\tau$~Herculid meteors recorded by video cameras (J2000.0).}             
\centering          
\begin{tabular}{l c c c c} 
\hline
Meteor code		&  Time [UT]	&	$\alpha_{G}$ [$^{\circ}$]	&	$\delta_{G}$ [$^{\circ}$]	&	$V_{G}$ [km s$^{-1}$]	\\
\hline
22530012        &  21:15:42.4	&   209.29$\pm$0.11		&   27.96$\pm$0.11	&	11.76$\pm$0.18			\\  
22530021        &  21:50:05.3   &	206.5$\pm$0.2		&	26.75$\pm$0.19	&	11.8$\pm$0.4			\\  
22530043        &  22:56:08.3	&   206.63$\pm$0.16		&	26.56$\pm$0.15	&	11.8$\pm$0.2			\\ 
22530044        &  22:58:40.4	&   209.49$\pm$0.10		&	28.7$\pm$0.2	&	12.74$\pm$0.14			\\  
22530045        &  22:59:24.1	&   207.4$\pm$0.3		&   26.7$\pm$0.3	&	11.5$\pm$0.6			\\ 
22530049        &  23:07:27.2	&   208.5$\pm$0.2		&	27.9$\pm$0.2	&	11.2$\pm$0.3			\\  
22530050		&  23:11:07.4	&   207.34$\pm$0.11		&	25.9$\pm$0.2	&	11.61$\pm$0.1			\\  
22530052		&  23:12:04.4	&   209.68$\pm$0.19		&	28.53$\pm$0.19	&	11.8$\pm$0.3			\\  
22530054		&  23:14:04.3	&   209.66$\pm$0.14		&	28.03$\pm$0.08	&	11.76$\pm$0.10			\\
22530056		&  23:18:06.4	&   209.72$\pm$0.16		&	26.8$\pm$0.2	&	10.7$\pm$0.2			\\  
22530061		&  23:25:01.4	&	208.52$\pm$0.12		&	28.29$\pm$0.10	&	11.01$\pm$0.15			\\ 
22530065		&  23:32:52.3	&   207.74$\pm$0.12		&   26.61$\pm$0.11	&	11.51$\pm$0.17			\\  
22530068        &  23:33:26.3 	&   209.04$\pm$0.10		&   28.04$\pm$0.09	&   11.66$\pm$0.12			\\
22530077        &  23:47:32.3	&   209.09$\pm$0.16		&   28.23$\pm$0.13  &   12.06$\pm$0.18			\\
22530080        &  23:52:45.3 	&   209.46$\pm$0.10		&   28.30$\pm$0.07  &   12.02$\pm$0.09			\\
22530086        &  00:02:50.7   &   205.6$\pm$0.6  		&   26.0$\pm$0.3    &   11.0$\pm$0.2			\\
22530092        &  00:09:40.5   &   208.86$\pm$0.06  	&   28.40$\pm$0.11  &  	12.08$\pm$0.05			\\
22530104        &  00:28:19.3   &   209.19$\pm$0.16  	&   28.01$\pm$0.12  &  	12.12$\pm$0.12			\\
22530120        &  00:47:45.2   &   209.3$\pm$0.2  		&   28.0$\pm$0.2    &  	11.6$\pm$0.2			\\
22530139        &  01:19:02.4   &   208.70$\pm$0.15  	&   27.51$\pm$0.14  &  	11.35$\pm$0.14 			\\  
22530140        &  01:19:20.2   &   207.1$\pm$0.2  		&   26.5$\pm$0.2    &  	11.0$\pm$0.2 			\\  
22530143        &  01:22:15.9   &   209.72$\pm$0.06  	&   28.59$\pm$0.06  &  	12.18$\pm$0.06 			\\  
22530146        &  01:35:48.6   &   209.34$\pm$0.10  	&   28.20$\pm$0.09  &  	11.98$\pm$0.111 		\\  
22530501        &  20:39:12.5   &   209.3$\pm$0.4  		&   27.54$\pm$0.15  &  	11.0$\pm$0.2 			\\  
22530502        &  20:39:27.5   &   209.5$\pm$0.2  		&   28.02$\pm$0.15  &  	11.9$\pm$0.3 			\\  
22530504        &  20:46:54.3   &   205.72$\pm$0.14  	&   24.7$\pm$0.6    &  	11.0$\pm$0.6 			\\  
22530513        &  21:18:54.6   &   208.67$\pm$0.15  	&   28.21$\pm$0.15  &  	11.3$\pm$0.3 			\\  
22530524        &  21:48:36.2   &   209.5$\pm$0.2  		&   27.51$\pm$0.19  &  	12.3$\pm$0.5 			\\  
22530530        &  22:02:40.5   &   206.1$\pm$0.5  		&   24.8$\pm$0.55   &  	11.14$\pm$0.18 			\\  
22530538        &  22:12:44.7   &   210.07$\pm$0.16     &   27.6$\pm$0.2   	&   12.5$\pm$0.5			\\  
22530546        &  22:25:14.3   &   208.0$\pm$0.2  		&   27.79$\pm$0.16  &   12.7$\pm$0.3			\\  
22530547        &  22:25:24.6   &   209.25$\pm$0.08  	&   27.97$\pm$0.08  &   11.77$\pm$0.17			\\  
22530553        &  22:31:49.2   &   209.01$\pm$0.14  	&   27.22$\pm$0.16  &   12.1$\pm$0.3			\\  
22530556        &  22:34:43.2   &   208.3$\pm$0.3  		&   27.9$\pm$0.3   	&   11.6$\pm$0.5			\\  
22530562        &  22:40:24.3   &   208.49$\pm$0.15  	&   25.85$\pm$0.15  &   11.1$\pm$0.3			\\  
22530569        &  22:56:27.2   &   209.0$\pm$0.3  		&   27.6$\pm$0.3   	&   11.3$\pm$0.6			\\  
22530576        &  23:12:12.4   &   209.4$\pm$0.2  		&   27.0$\pm$0.3   	&   11.1$\pm$0.3			\\  
22530577        &  23:12:40.5   &   209.5$\pm$0.2  		&   27.7$\pm$0.2   	&   12.2$\pm$0.4			\\  
22530579        &  23:13:20.6   &   210.4$\pm$0.3  		&   27.7$\pm$0.2   	&   12.5$\pm$0.2			\\  
22530584        &  23:22:28.4   &   209.3$\pm$0.2  		&   26.7$\pm$0.2   	&   11.7$\pm$0.3			\\  
22530594        &  23:39:51.4   &   208.78$\pm$0.18  	&   28.02$\pm$0.15  &   11.0$\pm$0.2			\\  
22530597        &  23:49:46.5   &   209.76$\pm$0.16  	&   28.72$\pm$0.18  &   11.9$\pm$0.2			\\  
22530599        &  23:50:04.4   &   211.5$\pm$0.4  		&   26.3$\pm$0.3  	&   12.3$\pm$0.6			\\  
22530600        &  23:50:25.3   &   208.0$\pm$0.2  		&   28.0$\pm$0.2    &   12.0$\pm$0.4 			\\  
22530604        &  23:55:26.4   &   208.7$\pm$0.3  		&   26.2$\pm$0.2    &   11.7$\pm$0.3 			\\  
22530611        &  00:60:52.6   &   208.23$\pm$0.15  	&   26.71$\pm$0.13  &   10.90$\pm$0.18 			\\  
22530612        &  00:10:59.4   &   210.6$\pm$0.3  		&   28.2$\pm$0.2    &   13.5$\pm$0.2 			\\  
22530620        &  00:28:48.3   &   206.6$\pm$0.5  		&   27.0$\pm$0.3    &   10.2$\pm$0.4 			\\  
22530622        &  00:31:44.2   &   209.30$\pm$0.18  	&   26.77$\pm$0.12  &   11.94$\pm$0.18 			\\  
22530623        &  00:32:58.4   &   208.7$\pm$0.4  		&   27.5$\pm$0.3    &   11.2$\pm$0.5 			\\  
22530624        &  00:33:58.3   &   207.7$\pm$0.4  		&   25.3$\pm$0.3    &   10.6$\pm$0.4 			\\  
22530626        &  00:49:25.4   &   206.6$\pm$0.3  		&   24.81$\pm$0.18  &   10.54$\pm$0.17 			\\  
22530628        &  00:51:11.5   &   209.31$\pm$0.17  	&   27.87$\pm$0.14  &   11.9$\pm$0.2 			\\  
22530638        &  01:08:33.4   &   209.9$\pm$0.5  		&   27.3$\pm$0.4    &   11.5$\pm$0.5 			\\  
22530642        &  01:19:29.3   &   208.6$\pm$0.3  		&   27.0$\pm$0.2    &   10.9$\pm$0.2 			\\  
22530643        &  01:25:44.3   &   209.8$\pm$0.4  		&   28.1$\pm$0.3    &   11.9$\pm$0.4 			\\  
22530649        &  01:36:33.2   &   206.7$\pm$0.84  	&   25.4$\pm$0.7    &   11.2$\pm$0.7 			\\  
\hline
\end{tabular}
\end{table*}

\begin{table*}
\caption{Geocentric radiants of $\tau$~Herculid meteors recorded by photographic cameras (J2000.0).}             
\centering          
\begin{tabular}{l c c c c} 
\hline
Meteor code		&  Time [UT]	&	$\alpha_{G}$ [$^{\circ}$]	&	$\delta_{G}$ [$^{\circ}$]	&	$V_{G}$ [km s$^{-1}$]	\\
\hline
EN300522\_210631 &  21:00:32.02 	& 	209.00$\pm$0.11  	&   27.53$\pm$0.16  &   11.4$\pm$0.4 			\\  
EN300522\_211542 &  21:15:43.26 	& 	206.45$\pm$0.13  	&   25.68$\pm$0.13  &   11.5$\pm$0.2 			\\  
EN300522\_211854 &  21:18:55.45 	& 	208.85$\pm$0.04  	&   28.27$\pm$0.08  &   10.92$\pm$0.17 			\\  
EN300522\_213753 &  21:37:53.22     &   207.20$\pm$0.10     &   25.55$\pm$0.09  &   11.93$\pm$0.18			\\                   
EN300522\_214039 &  21:40:39.35 	& 	209.08$\pm$0.13  	&   27.68$\pm$0.13  &   11.26$\pm$0.15 			\\  
EN300522\_214616 &  21:46:16.99 	& 	208.02$\pm$0.05  	&   26.34$\pm$0.09  &   11.99$\pm$0.11 			\\  
EN300522\_214805 &  21:48:06.17  	& 	209.45$\pm$0.05  	&   27.93$\pm$0.08  &   11.79$\pm$0.14 			\\  
EN300522\_221454 &  22:14:54.21     &   208.61$\pm$0.17     &   27.5$\pm$0.2    &   10.4$\pm$0.3			\\                   
EN300522\_230822 &  23:08:23.09 	& 	209.02$\pm$0.04  	&   28.24$\pm$0.04  &   11.81$\pm$0.03 			\\  
EN300522\_231358 &  23:13:59.21 	& 	209.52$\pm$0.10  	&   28.03$\pm$0.04  &   12.00$\pm$0.06 			\\  
EN300522\_233047 &  23:30:47.29     &   209.34$\pm$0.11     &   28.79$\pm$0.13  &   12.05$\pm$0.16		    \\                   
EN300522\_233541 &  23:35:41.56     &   209.17$\pm$0.08     &   27.88$\pm$0.07  &   12.21$\pm$0.14		    \\                   
EN310522\_000525 &  00:05:25.40     &   205.9$\pm$0.3       &   24.4$\pm$0.3    &   10.1$\pm$0.3		    \\                   
EN310522\_002838 &  00:28:38.26 	& 	210.16$\pm$0.14  	&   28.59$\pm$0.10  &   12.72$\pm$0.19 			\\  
EN310522\_002928 &  00:29:29.17 	& 	206.60$\pm$0.09  	&   25.89$\pm$0.09  &   11.56$\pm$0.10 			\\  
EN310522\_003132 &  00:31:32.36     &   208.1$\pm$0.4       &   27.0$\pm$0.3    &   10.2$\pm$0.4		   \\                   
EN310522\_003143 &  00:31:43.50     &   208.0$\pm$0.2       &   27.24$\pm$0.17  &   11.1$\pm$0.3		   \\                   
EN310522\_004252 &  00:42:52.15     &   209.54$\pm$0.14     &   28.12$\pm$0.12  &   12.23$\pm$0.20         \\                   
EN310522\_011235 &  01:12:35.13 	& 	209.06$\pm$0.17  	&   27.83$\pm$0.14  &   12.00$\pm$0.10 			\\  
EN310522\_011503 &  01:15:03.19     &   208.2$\pm$0.3       &   26.7$\pm$0.2    &   12.0$\pm$0.3            \\                   
EN310522\_012448 &  01:24:48.37     &   209.5$\pm$0.3       &   28.33$\pm$0.20  &   12.2$\pm$0.3           \\                   
EN310522\_012841 &  01:28:40.84 	& 	209.06$\pm$0.14  	&   28.08$\pm$0.13  &   11.68$\pm$0.15 			\\  
EN310522\_013924 &  01:39:23.92 	& 	208.3$\pm$0.5  		&   28.3$\pm$0.2 	&   11.94$\pm$0.03 			\\  
\hline
\end{tabular}
\end{table*}

\begin{table*}
\caption{Heliocentric orbits of $\tau$~Herculid meteors recorded by video cameras (J2000.0).}             
\centering          
\begin{tabular}{l c c c c c c c} 
\hline
Meteor code		   &  		a [au]		 &		   e			  &			q [au] 			&	   i [$^{\circ}$]  &   $\omega$ [$^{\circ}$] & $\Omega$ [$^{\circ}$]	&  $T_{J}$	\\
\hline
22530012           &    2.91$\pm$0.10    &    0.660$\pm$0.012     &      0.9893$\pm$0.0003   &    10.84$\pm$0.15   &     200.04$\pm$0.07    &    69.1538$\pm$0.0002    &    2.89    \\       
22530021           &    3.2$\pm$0.3      &    0.69$\pm$0.03       &      0.9914$\pm$0.0007   &    10.2$\pm$0.3     &     198.89$\pm$0.12    &    69.1776$\pm$0.0005    &    2.74    \\       
22530043           &    3.19$\pm$0.14    &    0.689$\pm$0.014     &      0.9912$\pm$0.0004   &    10.19$\pm$0.18   &     199.01$\pm$0.10    &    69.2217$\pm$0.0003    &    2.75    \\       
22530044           &    3.58$\pm$0.12    &    0.724$\pm$0.009     &      0.9885$\pm$0.0004   &    11.73$\pm$0.13   &     199.86$\pm$0.12    &    69.2212$\pm$0.0002    &    2.57    \\       
22530045           &    2.9$\pm$0.3      &    0.66$\pm$0.04       &      0.9909$\pm$0.0011   &    10.1$\pm$0.5     &     199.37$\pm$0.16    &    69.2239$\pm$0.0008    &    2.87    \\       
22530049           &    2.67$\pm$0.13    &    0.629$\pm$0.018     &      0.9914$\pm$0.0006   &    10.3$\pm$0.3     &     199.48$\pm$0.11    &    69.2290$\pm$0.0004    &    3.04    \\       
22530050           &    3.01$\pm$0.06    &    0.672$\pm$0.007     &      0.9899$\pm$0.0003   &    10.00$\pm$0.10   &     199.70$\pm$0.11    &    69.2320$\pm$0.0002    &    2.84    \\       
22530052           &    2.87$\pm$0.16    &    0.66$\pm$0.02       &      0.9897$\pm$0.0006   &    11.0$\pm$0.3     &     199.96$\pm$0.10    &    69.2311$\pm$0.0004    &    2.91    \\       
22530054           &    2.88$\pm$0.05    &    0.657$\pm$0.006     &      0.9891$\pm$0.0003   &    10.90$\pm$0.09   &     200.18$\pm$0.08    &    69.2326$\pm$0.0001    &    2.91    \\       
22530056           &    2.39$\pm$0.08    &    0.586$\pm$0.014     &      0.9896$\pm$0.0005   &    9.8 $\pm$0.2     &     200.68$\pm$0.11    &    69.2369$\pm$0.0004    &    3.26    \\       
22530061           &    2.57$\pm$0.06    &    0.614$\pm$0.009     &      0.9922$\pm$0.0003   &    10.27$\pm$0.14   &     199.27$\pm$0.07    &    69.2408$\pm$0.0002    &    3.12    \\       
22530065           &    2.91$\pm$0.09    &    0.659$\pm$0.011     &      0.9904$\pm$0.0004   &    10.15$\pm$0.16   &     199.61$\pm$0.06    &    69.2462$\pm$0.0002    &    2.90    \\       
22530068           &    2.87$\pm$0.06    &    0.655$\pm$0.008     &      0.9901$\pm$0.0003   &    10.75$\pm$0.11   &     199.78$\pm$0.06    &    69.2457$\pm$0.0002    &    2.92    \\       
22530077           &    3.11$\pm$0.11    &    0.682$\pm$0.011     &      0.9896$\pm$0.0004   &    11.09$\pm$0.17   &     199.76$\pm$0.09    &    69.2546$\pm$0.0002    &    2.78    \\       
22530080           &    3.05$\pm$0.05    &    0.676$\pm$0.006     &      0.9893$\pm$0.0002   &    11.12$\pm$0.09   &     199.94$\pm$0.06    &    69.2580$\pm$0.0001    &    2.81    \\       
22530086           &    2.76$\pm$0.10    &    0.640$\pm$0.014     &      0.9934$\pm$0.0008   &    9.4 $\pm$0.2     &     198.5$\pm$0.4      &    69.2675$\pm$0.0004    &    2.99    \\       
22530092           &    3.14$\pm$0.03    &    0.685$\pm$0.003     &      0.9901$\pm$0.0002   &    11.11$\pm$0.05   &     199.53$\pm$0.06    &    69.2693$\pm$0.0001    &    2.77    \\       
22530104           &    3.15$\pm$0.08    &    0.686$\pm$0.008     &      0.9891$\pm$0.0003   &    11.09$\pm$0.12   &     199.91$\pm$0.10    &    69.2818$\pm$0.0002    &    2.76    \\       
22530120           &    2.82$\pm$0.11    &    0.649$\pm$0.014     &      0.9899$\pm$0.0005   &    10.7$\pm$0.2     &     199.92$\pm$0.12    &    69.2952$\pm$0.0003    &    2.95    \\       
22530139           &    2.74$\pm$0.06    &    0.639$\pm$0.008     &      0.9906$\pm$0.0003   &    10.37$\pm$0.14   &     199.73$\pm$0.08    &    69.3166$\pm$0.0002    &    3.00    \\       
22530140           &    2.69$\pm$0.10    &    0.631$\pm$0.014     &      0.9919$\pm$0.0005   &    9.7 $\pm$0.2     &     199.21$\pm$0.12    &    69.3179$\pm$0.0004    &    3.03    \\       
22530143           &    3.13$\pm$0.04    &    0.684$\pm$0.004     &      0.9891$\pm$0.0001   &    11.34$\pm$0.06   &     199.94$\pm$0.03    &    69.3174$\pm$0.0001    &    2.77    \\       
22530146           &    3.04$\pm$0.06    &    0.675$\pm$0.006     &      0.9895$\pm$0.0002   &    11.06$\pm$0.11   &     199.87$\pm$0.04    &    69.3268$\pm$0.0001    &    2.82    \\       
22530501           &    2.55$\pm$0.11    &    0.612$\pm$0.017     &      0.9902$\pm$0.0006   &    10.2$\pm$0.2     &     200.1$\pm$0.3      &    69.1304$\pm$0.0003    &    3.13    \\       
22530502           &    2.99$\pm$0.20    &    0.67$\pm$0.02       &      0.9888$\pm$0.0005   &    11.0$\pm$0.3     &     200.17$\pm$0.14    &    69.1294$\pm$0.0004    &    2.85    \\       
22530504           &    2.8$\pm$0.3      &    0.64$\pm$0.04       &      0.9916$\pm$0.0011   &    9.1 $\pm$0.5     &     199.2$\pm$0.2      &    69.1375$\pm$0.0010    &    2.97    \\       
22530513           &    2.70$\pm$0.12    &    0.633$\pm$0.017     &      0.9912$\pm$0.0005   &    10.5$\pm$0.2     &     199.49$\pm$0.10    &    69.1564$\pm$0.0003    &    3.02    \\       
22530524           &    3.26$\pm$0.3     &    0.70$\pm$0.03       &      0.9876$\pm$0.0009   &    11.1$\pm$0.4     &     200.42$\pm$0.13    &    69.1753$\pm$0.0005    &    2.71    \\       
22530530           &    2.83$\pm$0.10    &    0.650$\pm$0.013     &      0.9911$\pm$0.0009   &    9.29 $\pm$0.19   &     199.4$\pm$0.4      &    69.1876$\pm$0.0004    &    2.94    \\       
22530538           &    3.3$\pm$0.4      &    0.71$\pm$0.03       &      0.9866$\pm$0.0009   &    11.4$\pm$0.4     &     200.73$\pm$0.08    &    69.1911$\pm$0.0005    &    2.67    \\       
22530546           &    3.7$\pm$0.3      &    0.73$\pm$0.02       &      0.9894$\pm$0.0006   &    11.2$\pm$0.3     &     199.41$\pm$0.13    &    69.1996$\pm$0.0004    &    2.52    \\       
22530547           &    2.92$\pm$0.09    &    0.661$\pm$0.011     &      0.9895$\pm$0.0003   &    10.84$\pm$0.14   &     199.98$\pm$0.04    &    69.2002$\pm$0.0002    &    2.89    \\       
22530553           &    3.20$\pm$0.19    &    0.691$\pm$0.018     &      0.9882$\pm$0.0005   &    10.9$\pm$0.2     &     200.22$\pm$0.09    &    69.2044$\pm$0.0003    &    2.74    \\       
22530556           &    2.9$\pm$0.3      &    0.66$\pm$0.03       &      0.9909$\pm$0.0009   &    10.6$\pm$0.4     &     199.39$\pm$0.15    &    69.2068$\pm$0.0006    &    2.89    \\       
22530562           &    2.64$\pm$0.12    &    0.625$\pm$0.017     &      0.9894$\pm$0.0005   &    9.8 $\pm$0.2     &     200.38$\pm$0.08    &    69.2119$\pm$0.0004    &    3.07    \\       
22530569           &    2.7$\pm$0.3      &    0.63$\pm$0.04       &      0.9903$\pm$0.0011   &    10.4$\pm$0.5     &     199.90$\pm$0.16    &    69.2216$\pm$0.0008    &    3.03    \\       
22530576           &    2.60$\pm$0.13    &    0.62$\pm$0.02       &      0.9894$\pm$0.0007   &    10.2$\pm$0.3     &     200.40$\pm$0.15    &    69.2324$\pm$0.0005    &    3.09    \\       
22530577           &    3.2$\pm$0.3      &    0.69$\pm$0.03       &      0.9881$\pm$0.0009   &    11.1$\pm$0.4     &     200.30$\pm$0.10    &    69.2314$\pm$0.0005    &    2.75    \\       
22530579           &    3.29$\pm$0.17    &    0.670$\pm$0.016     &      0.9864$\pm$0.0006   &    11.4$\pm$0.2     &     200.86$\pm$0.18    &    69.2314$\pm$0.0003    &    2.70    \\       
22530584           &    2.87$\pm$0.18    &    0.66$\pm$0.02       &      0.9883$\pm$0.0007   &    10.5$\pm$0.3     &     200.51$\pm$0.12    &    69.2388$\pm$0.0005    &    2.91    \\       
22530594           &    2.57$\pm$0.10    &    0.614$\pm$0.015     &      0.9915$\pm$0.0005   &    10.3$\pm$0.2     &     199.56$\pm$0.08    &    69.2507$\pm$0.0004    &    3.12    \\       
22530597           &    2.97$\pm$0.13    &    0.667$\pm$0.014     &      0.9895$\pm$0.0005   &    11.2$\pm$0.2     &     199.92$\pm$0.10    &    69.2559$\pm$0.0003    &    2.86    \\       
22530599           &    3.1$\pm$0.4      &    0.68$\pm$0.04       &      0.9833$\pm$0.0015   &    11.1$\pm$0.6     &     202.21$\pm$0.19    &    69.2563$\pm$0.0008    &    2.80    \\       
22530600           &    3.2$\pm$0.2      &    0.69$\pm$0.02       &      0.9908$\pm$0.0007   &    10.9$\pm$0.3     &     199.19$\pm$0.12    &    69.2568$\pm$0.0005    &    2.75    \\       
22530604           &    2.97$\pm$0.18    &    0.67$\pm$0.02       &      0.9882$\pm$0.0007   &    10.3$\pm$0.3     &     200.43$\pm$0.14    &    69.2610$\pm$0.0004    &    2.86    \\       
22530611           &    2.56$\pm$0.07    &    0.613$\pm$0.011     &      0.9910$\pm$0.0004   &    9.81 $\pm$0.17   &     199.78$\pm$0.07    &    69.2694$\pm$0.0003    &    3.12    \\       
22530612           &    4.2$\pm$0.3      &    0.768$\pm$0.016     &      0.9851$\pm$0.0007   &    12.3$\pm$0.2     &     200.78$\pm$0.19    &    69.2688$\pm$0.0002    &    2.36    \\       
22530620           &    2.37$\pm$0.14    &    0.58$\pm$0.03       &      0.9944$\pm$0.0009   &    9.2 $\pm$0.4     &     198.5$\pm$0.2      &    69.2851$\pm$0.0008    &    3.28    \\       
22530622           &    3.05$\pm$0.10    &    0.676$\pm$0.011     &      0.9878$\pm$0.0004   &    10.69$\pm$0.17   &     200.53$\pm$0.10    &    69.2846$\pm$0.0002    &    2.81    \\       
22530623           &    2.7$\pm$0.2      &    0.63$\pm$0.03       &      0.9910$\pm$0.0010   &    10.3$\pm$0.4     &     199.71$\pm$0.15    &    69.2860$\pm$0.0007    &    3.04    \\       
22530624           &    2.49$\pm$0.15    &    0.60$\pm$0.02       &      0.9907$\pm$0.0009   &    9.2 $\pm$0.4     &     200.03$\pm$0.17    &    69.2885$\pm$0.0007    &    3.18    \\       
22530626           &    2.51$\pm$0.07    &    0.605$\pm$0.011     &      0.9912$\pm$0.0005   &    8.94 $\pm$0.17   &     199.54$\pm$0.14    &    69.2994$\pm$0.0004    &    3.16    \\       
22530628           &    3.02$\pm$0.11    &    0.672$\pm$0.012     &      0.9892$\pm$0.0004   &    10.94$\pm$0.19   &     200.02$\pm$0.08    &    69.2972$\pm$0.0003    &    2.83    \\       
22530638           &    2.7$\pm$0.2      &    0.64$\pm$0.03       &      0.9885$\pm$0.0012   &    10.5$\pm$0.5     &     200.62$\pm$0.18    &    69.3093$\pm$0.0008    &    3.01    \\       
22530642           &    2.55$\pm$0.09    &    0.612$\pm$0.013     &      0.9909$\pm$0.0006   &    9.9 $\pm$0.2     &     199.87$\pm$0.17    &    69.3176$\pm$0.0004    &    3.13    \\       
22530643           &    3.0$\pm$0.2      &    0.67$\pm$0.02       &      0.9889$\pm$0.0009   &    11.1$\pm$0.4     &     200.17$\pm$0.16    &    69.3201$\pm$0.0006    &    2.85    \\       
22530649           &    2.8$\pm$0.4      &    0.65$\pm$0.04       &      0.9913$\pm$0.0016   &    9.5 $\pm$0.7     &     199.4$\pm$0.3      &    69.3297$\pm$0.0013    &    2.96    \\  
\hline
\end{tabular}
\end{table*}

\begin{table*}
\caption{Heliocentric orbits of $\tau$~Herculid meteors recorded by photographic cameras (J2000.0).}             
\centering          
\begin{tabular}{l c c c c c c c} 
\hline
Meteor code		   &  		a [au]		 &		   e			  &			q [au] 			&	   i [$^{\circ}$]  &   $\omega$ [$^{\circ}$] & $\Omega$ [$^{\circ}$]	&  $T_{J}$	\\
\hline
EN300522\_210631    & 2.74$\pm$0.19       &    0.64$\pm$0.02       & 	0.9899$\pm$0.0006   & 	10.5$\pm$0.3	   &   	200.02$\pm$0.06  	&  69.1482$\pm$0.0005  	&   3.00   \\                 
EN300522\_211542    & 2.99$\pm$0.15       &    0.669$\pm$0.017     & 	0.9910$\pm$0.0004   & 	9.74 $\pm$0.19	   &   	199.28$\pm$0.08  	&  69.1555$\pm$0.0003  	&   2.85   \\                 
EN300522\_211854    & 2.51$\pm$0.07       &    0.605$\pm$0.011     & 	0.9918$\pm$0.0003   & 	10.24$\pm$0.15	   &   	199.53$\pm$0.02  	&  69.1568$\pm$0.0002  	&   3.16   \\                 
EN300522\_213753    & 3.25$\pm$0.13       &    0.695$\pm$0.013     &    0.9891$\pm$0.0003   &   10.11$\pm$0.14     &    199.84$\pm$0.06     &  69.1697$\pm$0.0002   &   2.72   \\              
EN300522\_214039    & 2.66$\pm$0.07       &    0.628$\pm$0.010     & 	0.9902$\pm$0.0003   & 	10.39$\pm$0.13	   &   	199.97$\pm$0.09  	&  69.1711$\pm$0.0002  	&   3.01   \\                 
EN300522\_214616    & 3.20$\pm$0.08       &    0.691$\pm$0.007     & 	0.9888$\pm$0.0002   & 	10.44$\pm$0.09	   &   	200.00$\pm$0.04  	&  69.1747$\pm$0.0001  	&   2.74   \\                 
EN300522\_214805    & 2.91$\pm$0.08       &    0.661$\pm$0.009     & 	0.9891$\pm$0.0003   & 	10.88$\pm$0.12	   &   	200.14$\pm$0.03  	&  69.1753$\pm$0.0002  	&   2.89   \\                 
EN300522\_221454    & 2.32$\pm$0.09       &    0.572$\pm$0.017     &    0.9923$\pm$0.0006   &   9.6$\pm$0.3        &    199.65$\pm$0.12     &  69.1951$\pm$0.0005   &   3.33   \\              
EN300522\_230822    & 2.96$\pm$0.02       &    0.665$\pm$0.002     & 	0.99007$\pm$0.00009 & 	10.90$\pm$0.03	   &   	199.70$\pm$0.03  	&  69.2288$\pm$0.0000  	&   2.87   \\                 
EN300522\_231358    & 3.04$\pm$0.03       &    0.674$\pm$0.004     & 	0.98886$\pm$0.00017 & 	11.05$\pm$0.05	   &   	200.12$\pm$0.06  	&  69.2323$\pm$0.0001  	&   2.82   \\                 
EN300522\_233047    & 3.07$\pm$0.10       &    0.677$\pm$0.010     &    0.9899$\pm$0.0003   &   11.24$\pm$0.15     &    199.65$\pm$0.07     &  69.2432$\pm$0.0002   &   2.80   \\              
EN300522\_233541    & 3.21$\pm$0.09       &    0.692$\pm$0.009     &    0.9888$\pm$0.0003   &   11.12$\pm$0.12     &    199.99$\pm$0.04     &  69.2467$\pm$0.0002   &   2.73   \\              
EN310522\_000525    & 2.37$\pm$0.11       &    0.58$\pm$0.02       &    0.9930$\pm$0.0007   &   8.5$\pm$0.3        &    199.21$\pm$0.13     &  69.2711$\pm$0.0007   &   3.28   \\              
EN310522\_002838    & 3.50$\pm$0.14       &    0.717$\pm$0.012     & 	0.9875$\pm$0.0004   & 	11.79$\pm$0.17	   &   	200.29$\pm$0.06  	&  69.2811$\pm$0.0002  	&   2.61   \\                 
EN310522\_002928    & 3.04$\pm$0.06       &    0.674$\pm$0.006     & 	0.9911$\pm$0.0002   & 	9.86 $\pm$0.09	   &   	199.21$\pm$0.05  	&  69.2844$\pm$0.0002  	&   2.82   \\                 
EN310522\_003132    & 2.30$\pm$0.14       &    0.57$\pm$0.03       &    0.9928$\pm$0.0010   &   9.4$\pm$0.4        &    199.44$\pm$0.17     &  69.2866$\pm$0.0008   &   3.34   \\              
EN310522\_003143    & 2.68$\pm$0.11       &    0.629$\pm$0.016     &    0.9915$\pm$0.0005   &   10.1$\pm$0.2       &    199.40$\pm$0.08     &  69.2856$\pm$0.0004   &   3.04   \\              
EN310522\_004252    & 3.20$\pm$0.12       &    0.691$\pm$0.012     &    0.9886$\pm$0.0004   &   11.24$\pm$ 0.19    &    200.08$\pm$0.05     &  69.2913$\pm$0.0002   &   2.74   \\              
EN310522\_011235    & 3.08$\pm$0.06       &    0.679$\pm$0.006     & 	0.9893$\pm$0.0003   & 	10.94$\pm$0.10	   &   	199.88$\pm$0.11  	&  69.3115$\pm$0.0001  	&   2.80   \\                 
EN310522\_011503    & 3.2$\pm$0.2         &    0.69$\pm$0.02       &    0.9891$\pm$0.0007   &   10.6$\pm$0.3       &    199.86$\pm$0.10     &  69.3136$\pm$0.0005   &   2.73   \\              
EN310522\_012448    & 3.15$\pm$0.18       &    0.686$\pm$0.018     &    0.9890$\pm$0.0006   &   11.3$\pm$0.3       &    199.95$\pm$0.10     &  69.3192$\pm$0.0004   &   2.76   \\              
EN310522\_012841    & 2.88$\pm$0.08       &    0.657$\pm$0.009     & 	0.9902$\pm$0.0003   & 	10.77$\pm$0.15	   &   	199.72$\pm$0.06  	&  69.3224$\pm$0.0002  	&   2.91   \\                 
EN310522\_013924    & 3.10$\pm$0.04       &    0.680$\pm$0.004     & 	0.9909$\pm$0.0006   & 	10.91$\pm$0.09     &   	199.2$\pm$0.3    	&  69.3294$\pm$0.0001  	&   2.79   \\                 
\hline
\end{tabular}
\end{table*}

\end{appendix}

\end{document}